\documentclass[prx,twocolumn,superscriptaddress,longbibliography]{revtex4-1}
\usepackage{CJK}
\usepackage{amssymb,amsmath}
\usepackage{graphicx,color}
\usepackage{lipsum}
\usepackage[normalem]{ulem}
\usepackage{hyperref}
\usepackage{lineno}
\usepackage[CaptionAfterwards]{fltpage}
\usepackage{lipsum}

\usepackage{CJK}
\usepackage{amssymb,amsmath}
\usepackage{graphicx,color}
\usepackage{lipsum}
\usepackage[normalem]{ulem}
\usepackage{hyperref}
\usepackage{lineno}
\usepackage[CaptionAfterwards]{fltpage}
\usepackage{lipsum}

\begin{document}
\title{Distinct elastic properties and their origins in glasses and gels}

\author{Yinqiao Wang}
    \affiliation{Research Center for Advanced Science and Technology, The University of Tokyo, 4-6-1 Komaba, Meguro-ku, Tokyo 153-8904, Japan}

\author{Michio Tateno}
    \affiliation{Research Center for Advanced Science and Technology, The University of Tokyo, 4-6-1 Komaba, Meguro-ku, Tokyo 153-8904, Japan}
    
\author{Hajime Tanaka}
    \email[Email address: ]{tanaka@iis.u-tokyo.ac.jp}
    \affiliation{Research Center for Advanced Science and Technology, The University of Tokyo, 4-6-1 Komaba, Meguro-ku, Tokyo 153-8904, Japan}
    \affiliation{Department of Fundamental Engineering, Institute of Industrial Science, The University of Tokyo, 4-6-1 Komaba, Meguro-ku, Tokyo 153-8505, Japan}
    

\begin{abstract}
Glasses and gels, widely encountered amorphous solids with diverse industrial and everyday applications, share intriguing similarities such as rigidity without crystalline order and dynamic slowing down during aging. However, the underlying differences between these two fascinating materials have remained elusive. Here we uncover distinct elastic properties concerning observation and aging times in glasses and gels, while delving into the underlying mechanisms. In glasses, we observe a gradual decrease in the shear modulus, while the bulk modulus remains constant throughout the observation time. In contrast, gels exhibit a decrease in both the shear and bulk moduli over the observation time. Additionally, during aging, glasses exhibit a steady trend of stiffening, while gels demonstrate initial stiffening followed by softening. By unravelling the intricate relationship between structure, dynamics, and elasticity, we attribute these differences to mechanisms that minimize free energy: structural ordering in glasses and interface reduction in gels. Our work not only uncovers the distinct behaviors of glasses and gels but also sheds light on the origin and evolution of elasticity in non-equilibrium disordered solids, offering significant implications for the application and design of amorphous materials.
\end{abstract}

\maketitle

\section{Introduction}
Glasses, including metallic, polymer, colloidal, and granular glasses, are formed through rapid quenching or densification processes~\cite{berthier2011theoretical,wang2012elastic}. These materials are characterized by a homogeneous density field at a scale beyond individual particles. Their disordered yet homogeneous nature imparts unique mechanical properties distinct from crystals, leading to diverse and widespread applications. In contrast, low-density solids characterized by space-spanning particle networks emerge from the dynamical arrest of phase separation~\cite{zaccarelli07review,royall21review}, which are often called physical gels. These materials exhibit extraordinary soft elasticity, which is vital in various applications within biological systems, food products, pharmaceuticals, and cosmetics.

Here we consider glasses and gels of spherical particles interacting with isotropic potentials. 
Glasses with such interactions sterically self-organize their structure while gaining vibrational entropy at the expense of configurational entropy~\cite{tanaka2019revealing}, as widely recognized for hard spheres.
Colloidal gels, composed of particles with short-range attractions and interparticle bonds that can form and break reversibly, are widely studied as model systems to investigate the formation, aging, and yielding processes of physical gels~\cite{zaccarelli07review,royall21review,bonn2017yield}. The mechanisms responsible for gelation have been a subject of intense debate. One proposed mechanism suggests the dense phase evolving glass transition~\cite{lu08gelation,zia2014micro,testard2011influence,testard2014intermittent}, suggesting a connection between glasses and gels as members of the same class of amorphous solids. Conversely, another mechanism proposes the percolation of local rigid structure~\cite{whitaker19glassy_cluster,tsurusawa19isostatic,hsiao12isostatic,royall08arrest,zhang19rigidity_percolation,tsurusawa2023hierarchical}. 
In this mechanism, gels are formed through a process of hierarchical ordering driven by local `potential energy'~\cite{tsurusawa2023hierarchical}. This indicates a fundamental difference between gels and glasses, as the latter's formation is primarily driven by `entropy'~\cite{tanaka2019revealing}.

Apart from the ongoing debates about the formation of glasses and gels, there is still uncertainty about how their material properties compare and differ once they are formed. One fundamental material property of interest is the elastic modulus, particularly the shear modulus, denoted as $G$, which is linked to nonergodicity and thus crucial in distinguishing between liquids and solids. In crystals, the origin of $G$ is well-established, arising from broken translational symmetry~\cite{alexander1998amorphous}. However, in amorphous solids, such as glasses, gels, or jammed materials, the structural changes during the transition from a liquid-like to a solid-like state are minimal and do not involve translational ordering.

For athermal amorphous solids or inherent states of supercooled liquids, the shear modulus $G$ has been attributed to the mechanical balance of forces acting within the system~\cite{degiuli2018field,lemaitre2018stress,nampoothiri2020emergent,wang2020connecting}. In amorphous solids at finite temperature, the emergence of $G$ has been explained using a mean-field theory approach~\cite{yoshino2010emergence,yoshino2014shear,szamel2011emergence,parisi2010mean} or through transient mechanical force balance~\cite{yanagishima2017common,tong20rigidity}. In a practical context, the shear modulus $G$ has also been linked to the mean-squared displacements, rescaled either by density~\cite{ding2016universal} or by the infinite-frequency modulus~\cite{saw16modulus_calculation}, which quantify the available configurational space.

Glasses and gels, being far from thermodynamic equilibrium states, undergo an aging process where their properties slowly change over time. The dynamics of amorphous solids generally exhibit slowing down during aging, indicating a decrease of available configurational space, which is expected to increase the shear modulus $G$. Such an age-stiffening phenomenon is observed in glasses~\cite{greinert2006measurement, wang2012elastic}. However, for physical gels, many rheology experiments \cite{poon1999delayed, teece14delayed_collapse, bartlett12delayed_collapse, kamp2009universal, clarke21peak_modulus, fenton2023minimal} demonstrate a different behavior: while $G$ initially increases gradually during aging, it eventually decreases. This phenomenon is related to the delayed collapse of gels under the influence of gravity \cite{poon1999delayed, teece14delayed_collapse, bartlett12delayed_collapse, kamp2009universal, clarke21peak_modulus, fenton2023minimal}. Moreover, researchers have extensively investigated the connection between the elastic properties and the static structure of colloidal gels~\cite{zhang19rigidity_percolation,bouzid2018computing,bantawa2023hidden}. However, these studies often neglect the influence of thermal fluctuations, leaving the combined impact of both static structure and dynamics on the elasticity of amorphous materials, including glasses and gels, at finite temperatures unclear even at a fundamental level.

To address these critical issues, we conducted dynamic simulations on glass formers and colloidal gels with isotropic interactions. First, we observed distinct behaviors in the observation-time-dependent moduli of glasses and gels. Specifically, as the observation time increases, the shear modulus $G$ of glasses exhibited a decrease while the bulk modulus $K$ remained constant; however, both $G$ and $K$ of gels showed decreasing trends. Then, selecting the plateau modulus as a meaningful measure of elasticity, we monitored its time evolution during the aging process of both systems.
In glasses, we observed an increase in $G$ with a slight decrease in $K$ during aging. In contrast, for gels, we observed a gradual increase in both $G$ and $K$, followed by a subsequent decrease. By comparing these results to the behaviors of the corresponding elastic moduli $G^{\rm IS}$ and $K^{\rm IS}$ of the inherent states, which were obtained by minimizing the corresponding age-dependent structure at finite temperature, we found that the shear modulus at finite temperature, $G$, is primarily influenced by the shear modulus at inherent states, $G^{\rm IS}$, and the vibrational mean-squared displacement. This connection also holds true for the bulk modulus of gels but not for the bulk modulus of glasses, as it is mainly controlled by the volumetric constraint. Furthermore, we revealed two distinct aging mechanisms in glasses and gels. In glasses, the decrease in bulk free energy leads to structural ordering, resulting in a slowing down of dynamics and an increase in $G$, while $G^{\rm IS}$ remained relatively constant. On the other hand, in gels, the decrease in interfacial free energy promotes the reduction of the interface area of a network, leading to an increase in particle-scale connectivity ($Z$) and a decrease in network-scale connectivity ($N_{\rm loop}$). The former slows down dynamics and increases the elasticity of the inherent state, while the latter significantly reduces it, ultimately resulting in a peak in elastic moduli ($G$ and $K$) at finite temperatures for gels. These findings shed light on the complex and contrasting mechanical behaviors of glasses and gels during their aging processes, and highlight the importance of considering both static and dynamic factors in understanding the elasticity of amorphous materials at finite temperatures.

\section{Results and discussion}

\begin{figure*}
	\centerline{\includegraphics[width = 18 cm]{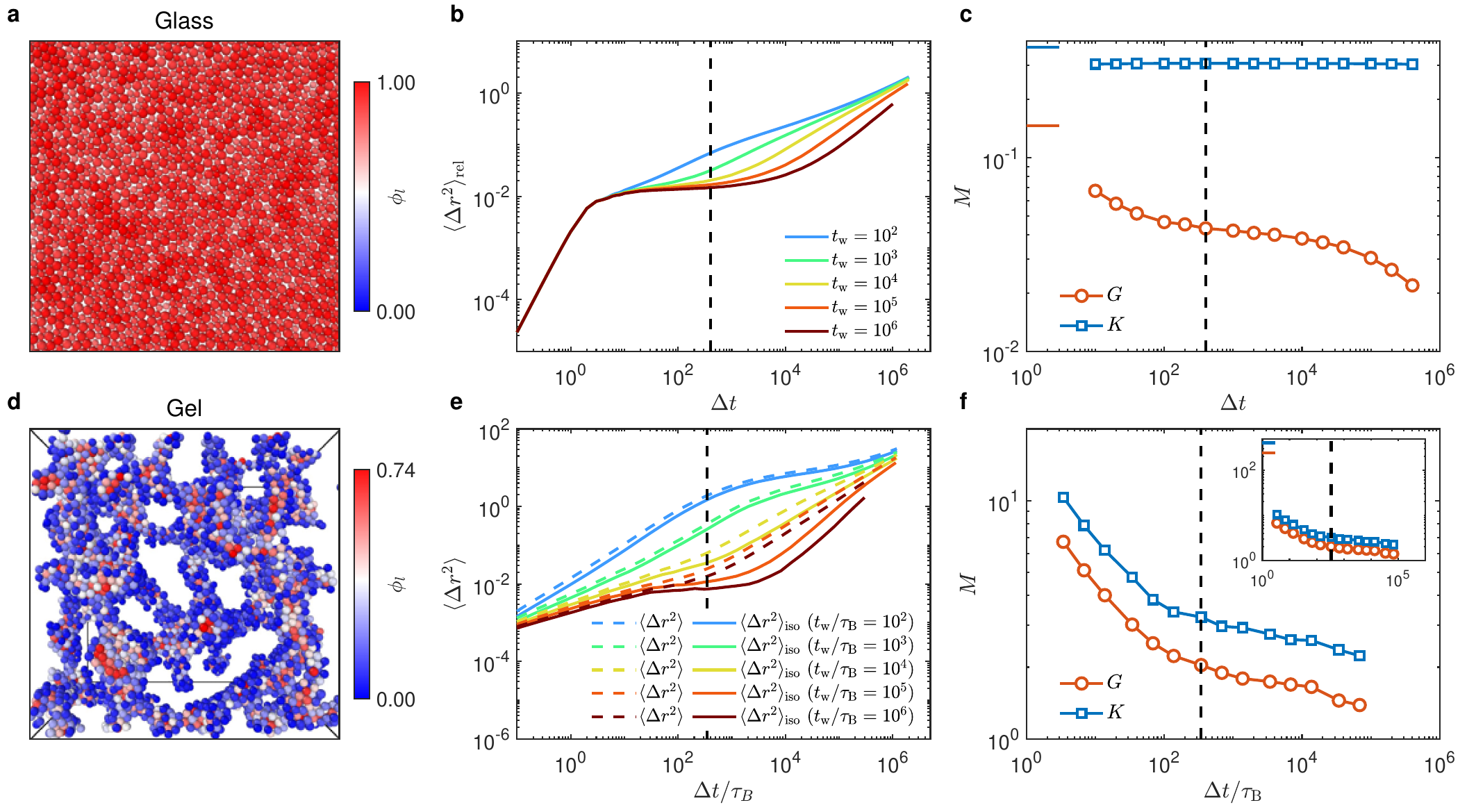}}
	\caption{aging dynamics and observation-time-dependent elasticity in glasses and gels. 
    Illustration of the typical structure of simple glass formers (a) and colloidal gels (d), where particles are colour-coded based on their local volume fraction $\phi_l$, representing the ratio of particle volume to the Voronoi volume.
    (b) Cage-relative mean-squared displacement (MSD) $\langle \Delta r^2\rangle_{\rm rel}$ versus observation time $\Delta t$ for different ages $t_{\rm w}$ of glasses.
    (e) MSD of all particles $\langle \Delta r^2\rangle$ and only isostatic particles $\langle \Delta r^2\rangle_{\rm iso}$ versus $\Delta t$ for different $t_{\rm w}$ of gels.
    (c) Observation-time-dependent shear modulus $G$ and bulk modulus $K$ of glasses. The short horizontal lines indicate the affine moduli $G_{\rm A}$ and $K_{\rm A}$.
    (f) Observation-time-dependent shear modulus $G$ and bulk modulus $K$ of gels. Inset: the same data as the main panel with the vertical axis zoomed out to show the affine moduli $G_{\rm A}$ and $K_{\rm A}$, indicated by the short horizontal lines. Here $\tau_{\rm B}$ denotes the Brownian time.
    }
	\label{fig:figure1}
\end{figure*}

We conducted numerical simulations on two distinct systems. The first one involved molecular dynamics simulations of a 2D binary mixture of particles with harmonic potentials, which is a well-studied glass-forming system~\cite{tong2018revealing,tong2019structural,tong20rigidity}. It is worth mentioning that this system exhibits consistent behaviors both in 2D and 3D~\cite{tong2018revealing,tong2019structural}. We selected the 2D system to allow for extended observation and aging times.
The second one involved Langevin dynamics simulations of a 3D polydisperse mixture of particles with Morse potentials~\cite{griffiths17low_density,zia2014micro}. This system represents a colloidal gel system that closely emulates the conditions observed in experimental colloid-polymer systems~\cite{tateno22tetra,tsurusawa19isostatic,tsurusawa2023hierarchical}.
In both systems, we initially equilibrated the particles at high-temperature liquid states. Subsequently, we instantaneously quenched the systems from temperatures well above the glass transition temperature $T_{\rm g}$ or the demixing critical temperature $T_{\rm c}$ to temperatures below these thresholds. The time elapsed from the quench is referred to as the waiting time or aging time, denoted as $t_{\rm w}$. Further simulation details can be found in the Methods section.
Figure~\ref{fig:figure1}a and d display typical structure of glasses and gels, respectively. Glasses exhibit a homogeneous density field beyond a scale above a few particle diameters (Fig.~\ref{fig:figure1}a). Conversely, gels exhibit a highly inhomogeneous density field and possess a large amount of interface area between the colloid-rich and colloid-poor phases (Fig.~\ref{fig:figure1}d).

Figure~\ref{fig:figure1}b and e, along with Fig.~\ref{fig:Sfigure1}, illustrate the age-dependent dynamics of both glasses and gels, revealing a noticeable slowing down in both systems. Glasses exhibit a clear two-step relaxation behavior, and their mean-squared displacement (MSD) displays a plateau region, which represents the vibrational MSD, akin to the Debye-Waller factor~\cite{widmer2006free, widmer2006predicting}. Here, we use the cage-relative MSD denoted as $\langle \Delta r^2\rangle_{\rm rel}$ to eliminate long-wavelength Mermin-Wagner fluctuations characteristic of 2D systems~\cite{flenner2015fundamental,shiba2016unveiling,vivek2017long,illing2017mermin}. On the other hand, the MSD of gels shows a crossover from subdiffusion to normal diffusion, but it retains a plateau region when considering only locally rigid (isostatic) particles with a contact number $Z\geq6$. This observation indicates that 
the contribution of highly mobile particles around the interface~\cite{zia2014micro} masks the plateau in MSD. Here, it is worth mentioning that isostatic particles are associated with the emergence of elasticity in the gel network~\cite{tsurusawa19isostatic, hsiao12isostatic}.

\subsection{Observation-time-dependent elasticity in glasses and gels}

For determining the elasticity of materials, two methods are often utilized~\cite{mizuno2022computational, saw16modulus_calculation, kriuchevskyi2017shear, zia2014micro, bouzid2018computing,lemaitre06modulus}. The first method involves directly applying external strain and measuring its stress response. For example, one can obtain viscoelastic moduli from oscillatory shear experiments~\cite{zia2014micro, bouzid2018computing, bantawa2023hidden}. The second method is the fluctuation formulation based on linear response theory~\cite{saw16modulus_calculation, kriuchevskyi2017shear, mizuno2022computational,lemaitre06modulus}. This approach considers the total modulus, which is the difference between the affine and nonaffine moduli. The affine component represents the energy increase when particles conform to homogeneous or affine deformation, while the nonaffine component arises from particle displacements that deviate from affine deformation. In thermal systems, the nonaffine modulus can be described using stress fluctuations~\cite{tong20rigidity, saw16modulus_calculation, kriuchevskyi2017shear, mizuno2022computational}. On the other hand, for athermal systems, it can be directly expressed using the Hessian matrix~\cite{mizuno2022computational, lemaitre06modulus}. Detailed formulations of these methods are provided in the Methods section.

To determine the elastic moduli $G$ and $K$ of glasses, we utilize the stress fluctuation formalism. On the other hand, for gels, we employ oscillatory shear and compression to obtain the frequency-dependent shear and bulk moduli, respectively (see Fig.~\ref{fig:Sfigure3}). As depicted in Fig.~\ref{fig:figure1}c and f, the elasticity depends on the observation time and exhibits distinct trends between glasses and gels.

In glasses, the observation time $\Delta t$ represents the time scale used to analyse stress fluctuations. As $\Delta t$ increases, $G$ decreases from the infinite-frequency, or affine value $G_{\rm A}$, reaches a plateau, and eventually vanishes as particles become diffusive. This plateau is known as the plateau modulus, corresponding to the plateau region observed in the MSD, as shown in Fig. \ref{fig:figure1}b. In contrast, the bulk modulus $K$ of glasses remains constant after decreasing from $K_{\rm A}$ within a very short time at a fast $\beta$ timescale.

In gels, the observation time $\Delta t$ refers to the period of oscillatory deformation. Both $G$ and $K$ decrease from their respective affine values, reach plateau regions, and eventually vanish at large $\Delta t$.

The distinct observation-time dependence of $G$ and $K$ in glasses is attributed to the different responses of materials to volumetric and shear deformation. The bulk modulus $K$ arises from pressure or density changes upon volumetric deformation, but for a uniform material like glasses, these changes cannot be relaxed. As a result, $K$ remains relatively constant even as the observation time increases.
On the other hand, the shear modulus $G$ arises from stress anisotropy upon shear deformation. When the system undergoes particle rearrangements and loses the memory of the initial anisotropic state, it returns to an isotropic state, and $G$ eventually vanishes. Thus, the bulk modulus is mainly determined by volumetric constraints, such as density and mean contact number, while the shear modulus is controlled by configurational constraints, representing the configurational space the system can explore. 

In contrast, for gels, due to the existence of a large void space, even under compression, the gel network has space to dilate, allowing the increased pressure to be released through particle rearrangement, similar to the shear case. Therefore, the configurational constraints affect $G$ and $K$ of gels in the same manner, leading to the observed similar observation-time dependence for both moduli. More detailed discussions on the influence of configurational constraints on $G$ and $K$ in gels will be provided later.

\subsection{Age-dependent elasticity in glasses and gels}
\begin{figure*}
	\centerline{\includegraphics[width = 18 cm]{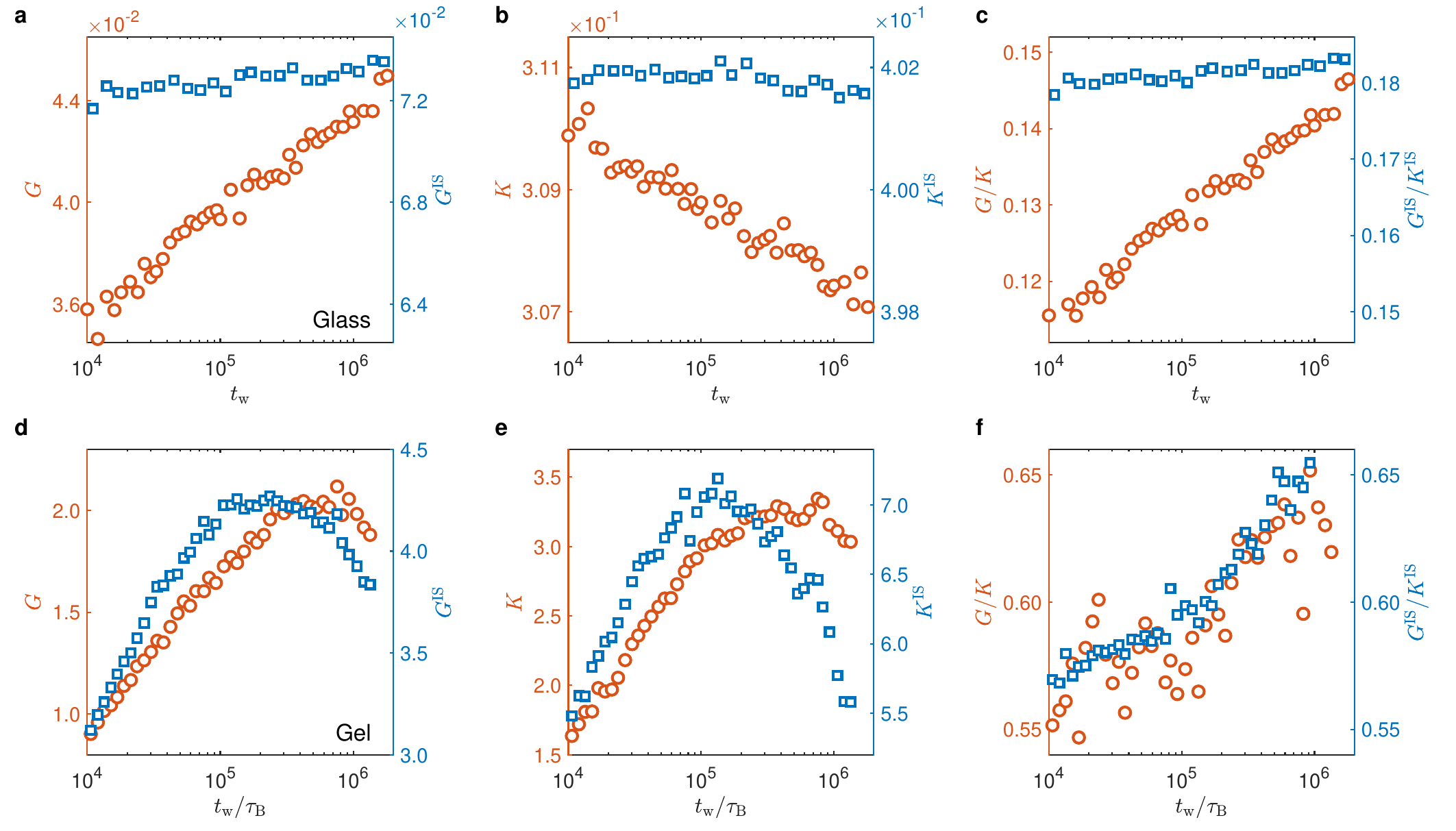}}
	\caption{Age-dependent elasticity in glasses and gels.
    The plateau moduli $G$, $K$, $G/K$ in finite temperature (red circles) and inherent moduli $G^{\rm IS}$, $K^{\rm IS}$, $G^{\rm IS}/K^{\rm IS}$ (blue squares) of glasses are plotted against waiting time $t_{\rm w}$ in (a), (b), and (c). The same elastic moduli of gels are displayed in (d), (e), and (f). The time $\Delta t$ used to measure these plateau moduli are marked by the vertical dashed lines in Fig.~\ref{fig:figure1}.}
	\label{fig:figure2}
\end{figure*}

To assess age-dependent elasticity, we employ the plateau modulus as a representative measurement. This modulus represents the timescale during which particle vibration is permitted while diffusion is prohibited. The plateau modulus characterizes how materials respond to small-amplitude external deformation, in which particles undergo nonaffine displacements while maintaining their neighbors. While the infinite-frequency modulus (also known as the affine modulus) can be easily determined by the mean number of contacts or bonds, and the zero-frequency modulus diminishes as particles become diffusive, the plateau modulus is a manifestation of the intrinsic disordered structure inherent in amorphous solids and holds profound physical significance. Here it is worth noting that the terminology for the plateau modulus may vary across different communities. In the context of metallic glasses, it is often referred to as the high- or infinite-frequency modulus~\cite{wang2012elastic, saw16modulus_calculation}, whereas for colloidal gels, it is commonly denoted as the low- or zero-frequency modulus~\cite{rocklin21elasticity}. This variation arises due to the relative position of measuring ranges of instruments and the characteristic vibrational frequency of materials.

For glasses, we choose the modulus at $\Delta t=400$ as the plateau timescale, which is indicated by dashed lines in Fig.~\ref{fig:figure1}b and c. Similar trends are observed for other timescales near the plateau region, as shown in Fig.~\ref{fig:Sfigure2}. As for gels, we select the period of oscillatory deformation as $\Delta t/\tau_{\rm B}=343$, marked by dashed lines in Fig.~\ref{fig:figure1}e and f.

To investigate the age-dependent static structure at finite temperatures, we derive the inherent states by eliminating thermal fluctuations using conjugate-gradient minimization. Then, we determine the elastic moduli of these inherent states using the Hessian matrix (see Methods), denoted as inherent elasticity or inherent moduli $G^{\rm IS}$ and $K^{\rm IS}$.

Figure~\ref{fig:figure2} presents the age-dependent thermal elasticity $M$ and inherent elasticity $M^{\rm IS}$ for both glasses and gels, revealing distinct trends.
In glasses, the shear modulus $G$ shows a monotonically increasing behavior over time, while it exhibits a peak in gels. Interestingly, in glasses, the bulk modulus $K$ experiences a slight decrease over time, in contrast to the trend observed for $G$. However, in gels, both $K$ and $G$ demonstrate similar trends.
One shared characteristic between glasses and gels is the increasing trend of the modulus ratio $G/K$, which indicates a decrease in Poisson's ratio.
Regarding the inherent moduli $G^{\rm IS}$ and $K^{\rm IS}$, they remain relatively stable for different ages in glasses, unlike the changes observed in the thermal state. This finding suggests that the variation in the thermal elasticity of glasses during aging is primarily influenced by thermal fluctuations. On the other hand, in gels, the inherent elasticity already exhibits a peak, indicating that the static structural evolution alone can lead to a peak in thermal elasticity. This implies that the thermal fluctuations have a more significant impact on the elasticity of glasses compared to gels. The contribution of the static structure and thermal fluctuations to the thermal elasticity will be further discussed in the following sections.

\subsection{Role of structure and dynamics on elasticity}

\begin{figure*}
	\centerline{\includegraphics[width = 18 cm]{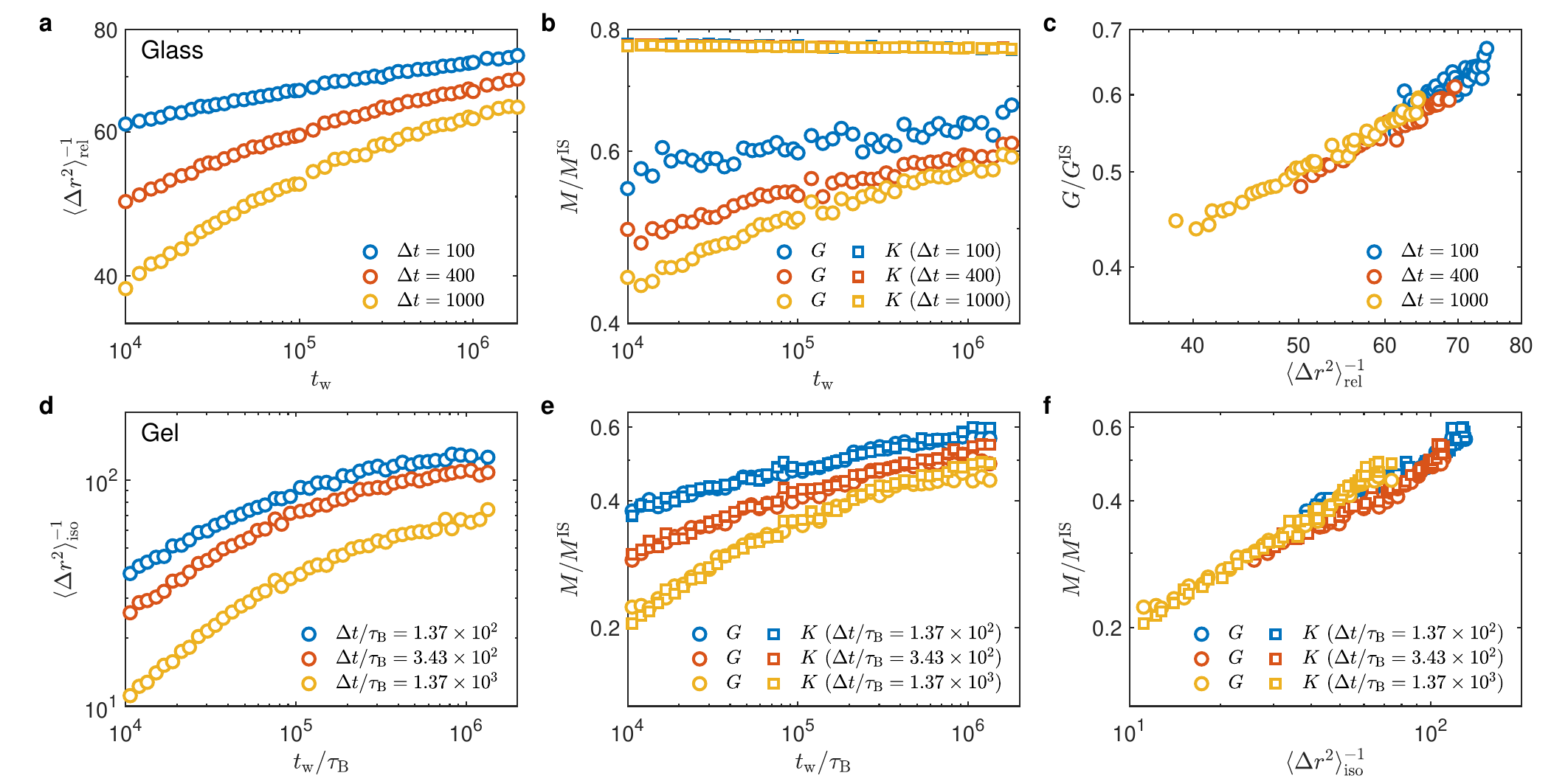}}
    \caption{Role of inherent elasticity and mean-squared displacement on thermal elasticity. (a) Inverse cage-relative MSD $\langle \Delta r^2\rangle^{-1}_{\rm rel}$ of glasses for different observation time $\Delta t$ plotted against waiting time $t_{\rm w}$. (b) $G/G^{\rm IS}$ and $K/K^{\rm IS}$ calculated using different statistic times $\Delta t$ for stress fluctuations in glasses. (c) $G/G^{\rm IS}$ versus $\langle \Delta r^2\rangle^{-1}_{\rm rel}$ for different $\Delta t$ in glasses. (d) Inverse MSD of isostatic particles ($Z\geq6$) $\langle \Delta r^2\rangle^{-1}_{\rm iso}$ of gels for different $\Delta t$ plotted against $t_{\rm w}$. (e) $G/G^{\rm IS}$ and $K/K^{\rm IS}$ of gels measured from different periods $\Delta t$ of oscillatory deformation. (f) $G/G^{\rm IS}$ and $K/K^{\rm IS}$ of gels versus $\langle \Delta r^2\rangle^{-1}_{\rm iso}$ for different $\Delta t$.}
	\label{fig:figure3}
\end{figure*}

As mentioned in the introduction, the shear modulus $G$ of the glass-liquid system follows a single proportional function of $\langle \Delta r^2\rangle^{-1}$ when rescaled by the infinite-frequency value $G_{\infty}=G_{\rm A}$ \cite{saw16modulus_calculation}. However, upon plotting $G/G_{\rm A}$ against $\langle \Delta r^2\rangle^{-1}$ in Fig.~\ref{fig:Sfigure4}a and b, we observe that while this approach works well for glasses, it fails for gels. In gels, $\langle \Delta r^2\rangle^{-1}$ increases monotonically, similar to glasses, as shown in Fig.~\ref{fig:figure3}a and d, which naturally leads to a failure in predicting the peak behavior of $G/G_{\rm A}$.

Notably, since the inherent elasticity $G^{\rm IS}$ of gels also exhibits a peak, we are inspired to rescale $G$ by $G^{\rm IS}$ instead of $G_{\rm A}$. As shown in Fig. \ref{fig:figure3}b and e, we plot the thermal elasticity rescaled by the inherent elasticity against the waiting time $t_{\rm w}$, which demonstrates a monotonic increase.

The bulk modulus of glasses appears insensitive to the configurational constraints, as discussed in the previous section. On the other hand, the shear modulus of glasses as well as the shear and bulk moduli of gels, exhibit similar trends with respect to $\langle \Delta r^2\rangle^{-1}$ when rescaled by the corresponding inherent moduli. To illustrate this, we plot $M/M^{\rm IS}$ as a function of $\langle \Delta r^2\rangle^{-1}$ in Fig.~\ref{fig:figure3}c and f, resulting in excellent collapses of all the curves. This rescaled elasticity shows a monotonic increase with aging, resembling the age-stiffening phenomenon observed in gels formed by van der Waals forces~\cite{bonacci2020contact}. However, it is important to note that, unlike that case, the stiffening in our case is not due to an increase in contact constraints, but rather it results from an increase of constraints in configurational space. 

As mentioned above, for 2D glasses, we employ the cage-relative MSD denoted as $\langle \Delta r^2\rangle_{\rm rel}$ to eliminate long-wavelength Mermin-Wagner fluctuations ~\cite{flenner2015fundamental,shiba2016unveiling,vivek2017long,illing2017mermin}. As shown in Fig. \ref{fig:Sfigure4}c, using the original MSD does not lead to the collapse of $G/G^{\rm IS}$. Similarly, in the case of 3D gels, employing the original MSD also fails to achieve a collapse of $M/M^{\rm IS}$, as illustrated in Fig.~\ref{fig:Sfigure4}d. This discrepancy can be attributed to the fast motion of low-coordination interface particles, which contributes little to the network's elasticity. Considering this point, we utilize the MSD of isostatic particles with a contact number $Z\geq 6$, denoted as $\langle \Delta r^2\rangle_{\rm iso}$, as shown in Fig.~\ref{fig:figure3}f. 

Furthermore, we observe that even the cage-relative MSD of isostatic particles for gels fails to result in a collapse of $M/M^{\rm IS}$, as shown in Fig.~\ref{fig:Sfigure4}e. This suggests that the vibrations in gels involve not only cage-relative motion but also collective motion of network strands at longer length-scales. Both of these types of motions contribute to the thermal elasticity of gels. This behavior is distinct from glasses, where collective motion only occurs at larger time scales, such as $\alpha$-relaxation time~\cite{tong2018revealing}. The presence of collective motion in gels further adds to the complexity of their mechanical properties.

Based on the collapse of $G/G^{\rm IS}$ versus $\langle \Delta r^2\rangle^{-1}$ for different observation times $\Delta t$, we can conclude that the shear modulus $G$ is universally determined by the inherent modulus $G^{\rm IS}$ and the vibrational MSD $\langle \Delta r^2\rangle^{-1}$. Therefore, the vibrational MSD effectively serves as a significant measure of configurational constraints in the material.
In contrast to glasses, the bulk modulus $K$ of gels shares the same configurational constraint with the shear modulus $G$, as demonstrated by the collapse of $K/K^{\rm IS}$ and $G/G^{\rm IS}$ in Fig.~\ref{fig:figure3}f, as well as the collapse of $G/K$ and $G^{\rm IS}/K^{\rm IS}$ in Fig.~\ref{fig:figure2}f. This suggests that the configurational constraints affecting the shear and bulk moduli in gels are strongly related and similarly influence their mechanical properties.

In the following sections, we will delve into the underlying mechanisms that drive the evolution of structure and dynamics during aging. We aim to elucidate how these factors control the evolution of the shear modulus $G$ and the bulk modulus $K$ in glasses and gels.

\subsection{Structural ordering in glasses and interface reduction in gels}

\begin{figure*}
	\centerline{\includegraphics[width = 18 cm]{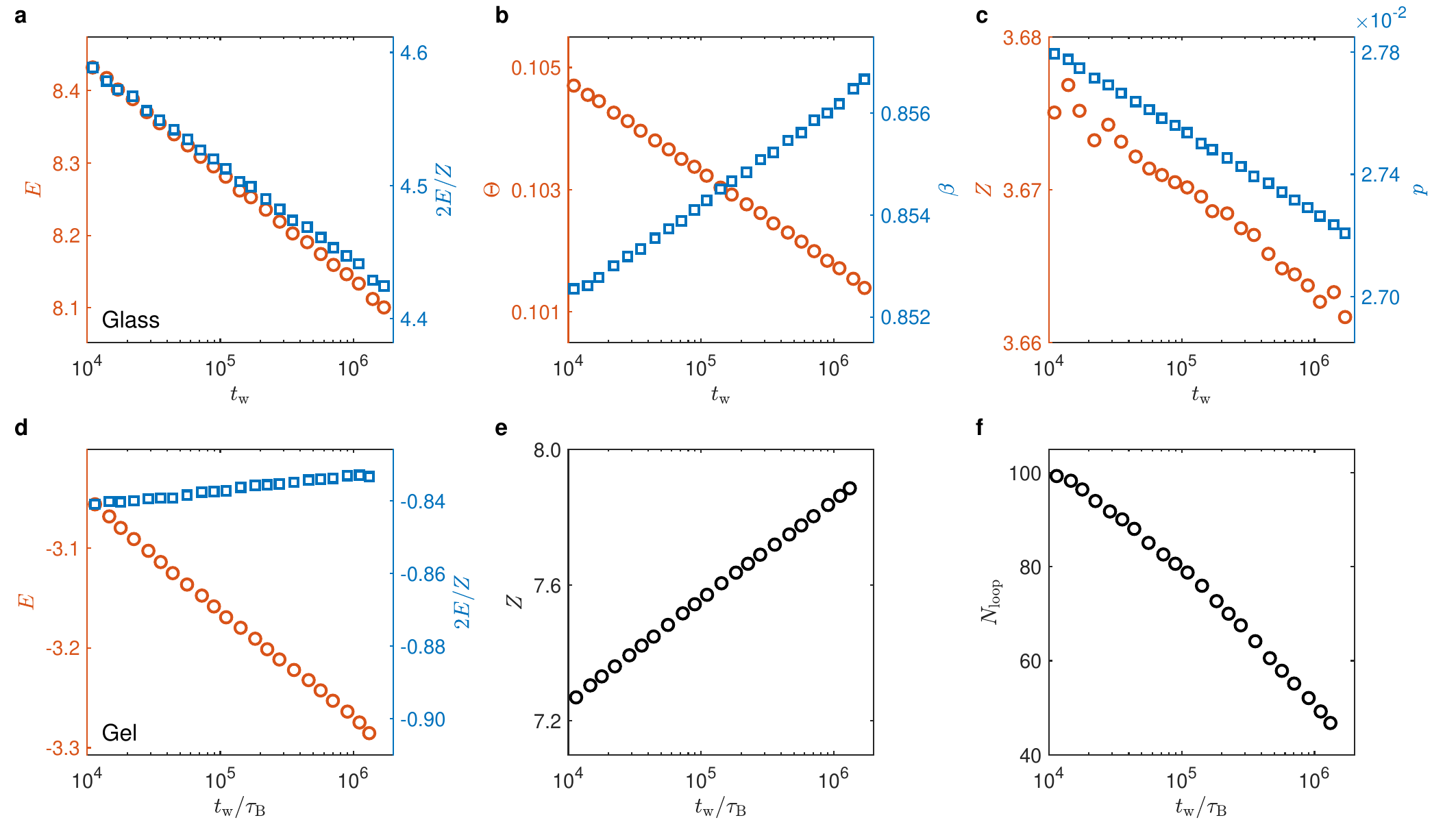}}
    \caption{Age-dependent potential energy and structure in glasses and gels.
    (a) Age-dependent potential energy per particle, $E$, and per contact, $2E/Z$, in glasses.
    (b) Orientational order parameter $\Theta$ and anisotropy parameter $\beta$ of Voronoi cells in glasses, where $\Theta=0$ indicates perfect order and $\beta=1$ indicates isotropic cells.
    (c) Mean contact number $Z$ and pressure $p$ in glasses.
    (d) Age-dependent potential energy per particle, $E$, and per contact, $2E/Z$, in gels.
    (e) Age-dependent mean contact number $Z$ in gels.
    (f) Age-dependent loop number $N_{\rm loop}$ in gels.}
	\label{fig:figure4}
\end{figure*}

\begin{figure*}[t!]
	\centerline{\includegraphics[width = 18 cm]{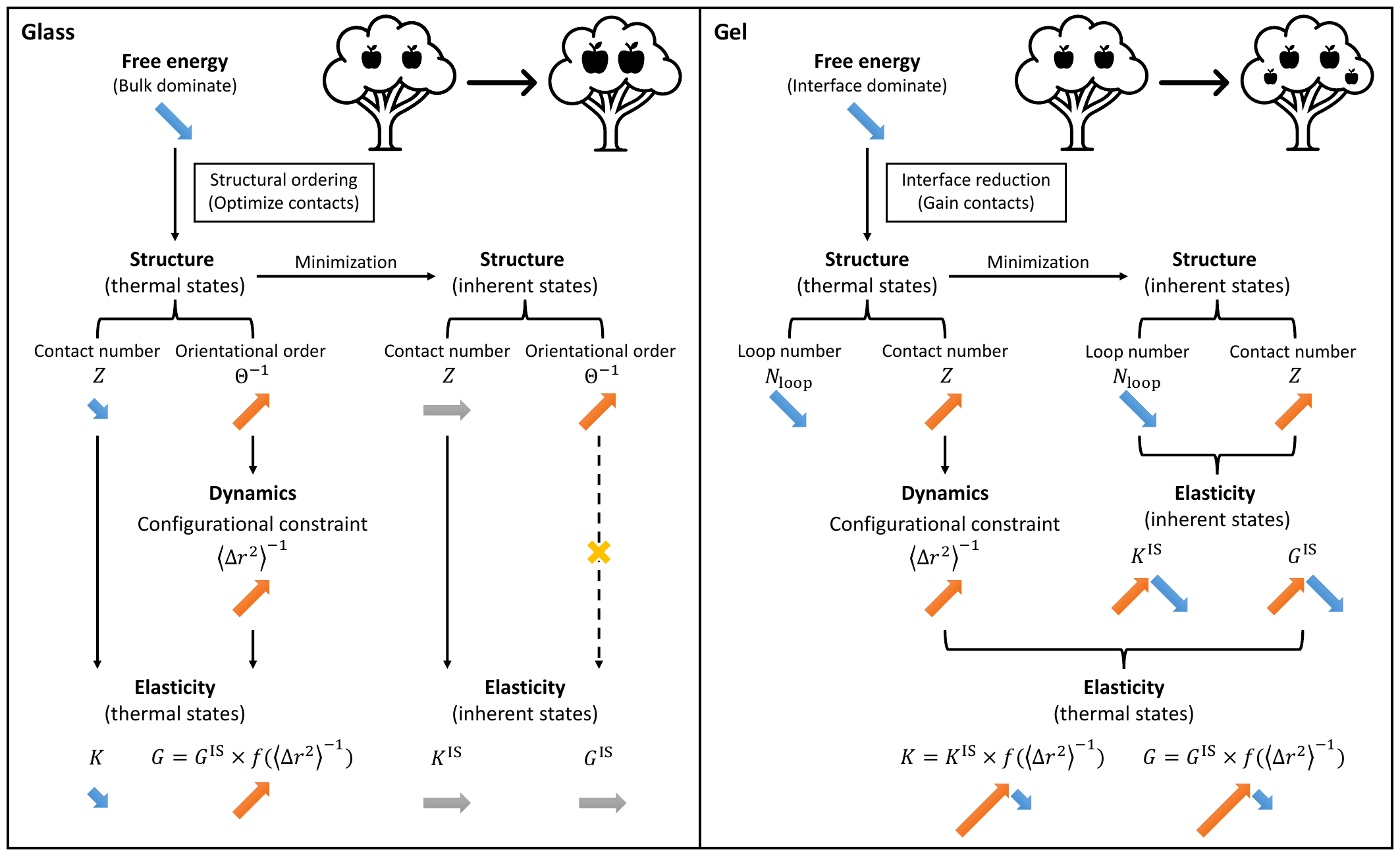}}
	\caption{Illustration of the origin of elasticity and distinct aging mechanisms in glasses and gels. The cartoon insets demonstrate two strategies for achieving a good harvest that corresponds to the decrease in potential energy in our problem. One strategy, illustrated for glasses, involves increasing the size of fruits (thus reducing the potential energy per contact) while keeping the number of fruits constant (the number of contacts). The other strategy, shown for gels, involves increasing the number of fruits (thereby increasing the number of contacts), even if the size of individual fruits (the potential energy per contact) decreases slightly on average. Additionally, we provide a schematic representation of the impacts of static structure and dynamics on elasticity for glasses (left) and gels (right).
	}
	\label{fig:figure5}
\end{figure*}

The fundamental driving force behind the aging process in non-equilibrium systems is their tendency to decrease free energy and approach a thermodynamic equilibrium state. Due to the different structure in glasses and gels, with uniform and non-uniform density, respectively, as shown in Fig. \ref{fig:figure1}a and d, the thermodynamic driving force is dominated by the bulk free energy for glasses and the interfacial free energy for gels. As measuring the absolute entropy poses challenges, we focus on the potential energy $E$, which holds significance for both soft-sphere glasses and attractive gels.

The potential energy $E$ can be expressed as the product of the mean contact number $Z$ and the potential energy per contact $2E/Z$. Here, we denote $E$ as the energy and $2E/Z$ as the contact energy. By utilising this decomposition, we identify two distinct ways to decrease the total energy in glasses and gels. As depicted in Fig. \ref{fig:figure4}a, in glasses, the decrease in contact energy contributes to over 90\% of the total energy decrease. However, in gels, the contact energy even increases by around 10\%, as shown in Fig. \ref{fig:figure4}d. This implies that the increase in the contact number contributes approximately 110\% to the overall decrease in the total energy. 

Based on these observations, we can draw the following conclusions. For glasses, with little change in the number of contacts, the structural evolution aims to decrease the contact energy or optimize contacts. In contrast, for gels, the system gains more contacts to decrease the energy, even though, on average, the contact energy increases slightly. Next, we will explore how these two ways of energy decrease drive the evolution of structure, dynamics, and elasticity in both glasses and gels in a coupled manner.

In glasses, the decrease in contact energy signifies an increase in packing capability, which can be quantified by an agnostic structural order parameter $\Theta$~\cite{tong2018revealing,tong2019structural,tanaka2019revealing}. This parameter measures the deviation of the local packing structure from the perfect packing structure and has been found to strongly correlate with the slowing dynamics in glass-forming liquids~\cite{tong2018revealing,tong2019structural,tanaka2019revealing}. As illustrated in Fig.~\ref{fig:figure4}b, $\Theta$ decreases logarithmically during the aging, indicating structural ordering, and leads to the slowing down of dynamics and an increase in the shear modulus $G$. Moreover, the decrease in $\Theta$ may also contribute to the increase in vibrational entropy in simple glass formers, at least in hard-sphere systems~\cite{tong2018revealing,tong2019structural,tanaka2019revealing}. This highlights the relationship between structural ordering, dynamics, and mechanical properties in glasses.

Recent studies have also explored the relationship between the slowing down of dynamics and a non-agnostic structural parameter~\cite{zhang2023anisotropic}, known as the cage anisotropy $\beta$. The cage anisotropy $\beta$ measures the ratio between the smallest and largest eigenvalues of the inertial moment of the Voronoi cell. As depicted in Fig.~\ref{fig:figure4}b, the Voronoi cells become more isotropic over time. The decrease in the Voronoi cell's anisotropy also contributes to the slowing down of dynamics, as particles tend to move maximally along the long axis of the Voronoi cell. 

In our simulations conducted under constant volume, an increase in structural ordering or packing capability leads to a slight decrease in volumetric constraints, such as pressure $p$ and mean contact number $Z$, as shown in Fig.~\ref{fig:figure4}c. Consequently, this leads to a slight decrease in $K$, as shown in Fig.~\ref{fig:figure2}b. This finding is not contradictory to the increase in $K$ observed in experiments~\cite{greinert2006measurement, wang2012elastic}. In experiments, structural ordering leads to densification under isobaric (constant-pressure) conditions, giving rise to the increase of $K$, which cannot occur in constant-volume simulations.

To disentangle the effects of static structure and thermal fluctuations on elasticity, we also analyse the structure and elasticity of glasses in inherent states. 
Interestingly, we observe that $K^{\rm IS}$ and $G^{\rm IS}$ remain almost constant, which is consistent with the minimal variation of $Z$ in inherent states, as shown in Fig.~\ref{fig:Sfigure5}b. This finding aligns with the existing understanding that both shear and bulk moduli in athermal jammed systems are determined by the contact number $Z$~\cite{ohern03jamming}. As presented in Fig.~\ref{fig:Sfigure5}a, the variations of $\Theta$ and $\beta$ in inherent states are comparable to those in thermal states.
These results indicate that the structural order parameters $\Theta$ or $\beta$ do not directly control the shear modulus, but instead determine the dynamics or the configurational constraints in the thermal state. These configurational constraints, in turn, influence the shear modulus $G$. 

Our results provide an explanation for the observed significant changes in the shear modulus $G$ with minimal alterations in the bulk modulus $K$ for different processing histories of glasses~\cite{cheng2009configurational}. The reason lies in the fact that the structural order, particularly the orientational order, can undergo substantial variations with minimal impact on volumetric constraints, such as density or mean contact number. 

In gels, the decrease in interfacial energy drives the increase in the contact number $Z$, as shown in Fig.~\ref{fig:figure4}e. As discussed earlier with respect to the vibrational MSD and the rescaled moduli $M/M^{\rm IS}$ in gels, the vibrational MSD encompasses both cage-relative local vibrations and vibrations on a strand-scale. It is evident that particles with more contacts exhibit lower mobility~\cite{zia2014micro}. Additionally, as the thickness of strands increases, the vibration of strands is expected to decrease. Thus, the increase in $Z$ leads to a monotonically increasing configurational constraint.
The peak modulus, as demonstrated, originates from static structure, and another impact of the increase in the contact number is the stiffening of the inherent structure. However, as the system decreases in interface area, thin strands are likely to rupture and be absorbed into the main part of the gel. Consequently, this process decreases the connectivity on a network scale, quantified by the loop number $N_{\rm loop}$ of the coarse-grained network in Fig.~\ref{fig:figure4}f, subsequently softening the inherent structure.
Both $Z$ and $N_{\rm loop}$ show a linear increase and decrease, respectively, against the logarithmic timescale. This suggests that the relative increase in $Z$ will become weaker over time, while the relative decrease in $N_{\rm loop}$ will become stronger, ultimately leading to the softening of gels.
The increase in the modulus ratio $G^{\rm IS}/K^{\rm IS}$ or $G/K$, as shown in Fig.~\ref{fig:figure2}f, indicates that the reduction in loop number has a stronger impact on volumetric deformation than on shear deformation.

\section{Outlook}

In summary, our study illustrates the intricate interplay among structure, dynamics, and elastic properties in non-equilibrium disordered systems by investigating two typical amorphous solids: simple glass formers and colloidal gels. Despite sharing many similarities as non-equilibrium amorphous solids, we reveal fundamental differences in their elastic properties between glasses and gels and elucidate the underlying mechanisms driving their respective behaviors. For clarity, we present a logic diagram in Fig.~\ref{fig:figure5}.

In the aging process of glasses, with uniform density fields, the dominant driving force is the bulk free energy. In contrast, gels exhibiting non-uniform density fields are governed by interfacial free energy. Consequently, the structural evolutions involved in decreasing free energy differ fundamentally between these two systems. 

In glasses, the decrease in bulk free energy is accomplished through structural ordering, which can be quantified by the orientational order parameter $\Theta^{-1}$. As a result, an increase in $\Theta^{-1}$ leads to the slowing down of dynamics, resulting in an elevation of the configurational constraint $\langle \Delta r^2 \rangle^{-1}$ and, ultimately, an increase in the shear modulus $G$. Furthermore, the structural ordering enhances the packing capability, resulting in a slight reduction in the volumetric constraint $Z$, which, in turn, causes a decrease in the bulk modulus $K$.

Upon investigating the inherent states, obtained by removing the thermal fluctuations of the corresponding age-dependent structures, we observe that $K^{\rm IS}$ remains constant due to the minimal variation in $Z$. Similarly, $G^{\rm IS}$ remains constant, indicating that the influence of the order parameter $\Theta^{-1}$ on the shear modulus $G$ is indirect, exerting its effect through the slowing down of dynamics, or configurational constraints. 
These findings shed light on the underlying mechanisms linking structural order, dynamics, and elastic properties in glasses, contributing to a deeper understanding of their complex behavior during aging.

In gels, the reduction of free energy is achieved by the reduction of interfacial area rather than structural ordering in the colloid-rich phase. Consequently, there is an increase in particle-scale connectivity (contact number $Z$) and a decrease in network-scale connectivity (loop number $N_{\rm loop}$). The increase in $Z$ elevates the configurational constraint $\langle \Delta r^2 \rangle^{-1}$ and increases the elasticity of inherent states. On the other hand, the decrease in $N_{\rm loop}$ tends to reduce the elasticity of inherent states. The competition between $Z$ and $N_{\rm loop}$ gives rise to a peak in the age-dependent $G^{\rm IS}$ and $K^{\rm IS}$.
The elasticity at finite temperature, which is a product of inherent elasticity $M^{\rm IS}$ (structure) and configurational constraint $\langle \Delta r^2 \rangle^{-1}$ (dynamics), also exhibits a peak during aging, although the decreasing trend at later stages is less pronounced compared to the elasticity in the inherent states. These findings highlight the intricate interplay between interface effects, structural connectivity, and dynamics in gels, leading to a complex evolution of their elastic properties during the aging process.

Notably, in glasses with uniform density, the bulk modulus is primarily affected by the bulk constraint. However, in gels with highly non-uniform density, the bulk modulus is also influenced by the configurational constraint. Consequently, an intriguing result emerges: at long observation times, when particle diffusion takes place, glasses behave like liquids with a finite bulk modulus and zero shear modulus, whereas gels resemble gases with zero bulk modulus and zero shear modulus.

Our work not only provides valuable insights into fundamental non-equilibrium physics but also has significant implications for materials science. It offers guidance on designing and manufacturing amorphous solids with desired elastic properties, by understanding the interplay between structural order, configurational constraints, and material response. These findings open up new possibilities for tailoring the mechanical behavior of amorphous materials, paving the way for innovative applications in various engineering and industrial fields.

\begin{acknowledgments}
H.T. acknowledges the support by the Grant-in-Aid for Specially Promoted Research (JSPS KAKENHI Grant No. JP20H05619) from the Japan Society for the Promotion of Science (JSPS). M.T. acknowledges the support from JSPS KAKENHI (Grant No. JP20K14424). Y.W. acknowledges the support from Shanghai Jiao Tong University via the scholarship for outstanding PhD graduates. 
\end{acknowledgments}


\section*{Methods}
\textbf{Simulation methods for simple glass formers.}
We conducted molecular dynamics simulations on a 2D binary mixture of particles interacting with harmonic potentials in a square box under the $NVT$ ensemble with a Nose-Hoover thermostat. The pairwise potential $u(r_{ij})$ between particles $i$ and $j$ is described by a harmonic function:
\begin{equation}
u(r_{ij}) = \frac{\epsilon}{2} \left(1 - \frac{r_{ij}}{d_{ij}}\right)^2, \quad r_{ij} < d_{ij},
\end{equation}
where $r_{ij}$ is the distance between particles $i$ and $j$, and $d_{ij}$ is the sum of their radii.

The system comprises $N=4096$ particles, with an equal number of large and small particles and a diameter ratio of 1.4. The diameter $d_{\rm s}$ of the small particles is set as the length unit, and both the small and large particles possess the same mass $m$. The energy, time, and temperature are scaled in units of $\epsilon$, $\sqrt{md_{\rm s}^2/\epsilon}$, and $10^{-3}\epsilon/k_{\rm B}$, respectively, where $k_{\rm B}$ represents the Boltzmann constant. The packing fraction of the system is fixed at $\phi=0.91$.

Prior studies have thoroughly investigated the glassy dynamics of these glass formers~\cite{tong2018revealing, tong2019structural, tong20rigidity}. The onset temperature $T_{\rm on}$ is approximately 2.1, the mode-coupling-theory (MCT) temperature $T_{\rm mct}$ is around 1.21, and the ideal glass transition temperature $T_0$ is roughly 0.63. In this study, we first equilibrated the system at $T=3.0$ and then rapidly quenched it to $T=1.0$ to observe the system's relaxation.

A total of 504 independent trajectories were performed for our investigation.

\textbf{Simulation methods for colloidal gels.}
We conducted Langevin dynamics simulations on a 3D polydisperse mixture of particles featuring short-range attractive potentials. The pairwise potential $u(r_{ij})$ between particles $i$ and $j$ is described by a Morse pair potential, given by:
\begin{equation}
u(r_{ij}) = \epsilon \exp{[\rho_M(d_{ij}-r_{ij})]}(\exp{[\rho_M(d_{ij}-r_{ij})]}-2),
\end{equation}
where the potential depth $\epsilon$ and the interaction range parameter $\rho_M$ can be independently controlled. Here, $r_{ij}$ represents the distance between particles $i$ and $j$, and $d_{ij}$ is the sum of their radii.

The system consists of $N=10000$ particles with a polydispersity of $\Delta/d=0.04$. The average particle diameter $d$ is taken as the length unit, and all particles have the same mass $m$. The energy, time, and temperature are scaled in units of $\epsilon$, $\sqrt{md^2/\epsilon}$, and $\epsilon/k_{\rm B}$, respectively, where $k_{\rm B}$ denotes the Boltzmann constant. The time step is set as $0.002$, and the damping factor is $\gamma=0.1$. The Brownian time $\tau_{\rm B}=(d/2)^2/(6D)$ is used to rescale the waiting time $t_{\rm w}$, where $D=\gamma k_{\rm B} T$ represents the self-diffusion constant for an isolated particle.

To align with typical PMMA colloid-polystyrene polymer experiments, we set the parameters as follows: packing fraction $\phi=0.12$, temperature $T=0.143$, and interaction range $\rho_M=48.63$. The range parameter $\rho_M$ is determined based on the second virial coefficients of the Asakura-Osawa (AO) potential~\cite{tateno19_fpd} with an attraction range of $\delta/d=0.13$, and the Morse potential is truncated and shifted at $r_{\rm cut}=1.13d_{ij}$.

We first equilibrate the system at a temperature of $T=1.0$, significantly higher than the demixing critical temperature of $T_c\sim0.3$~\cite{griffiths17low_density}. Subsequently, we instantaneously lower the temperature to the target value of $T=0.143$, thereby facilitating the aggregation of particles to form space-spanning networks.

A total of 23 independent trajectories were performed for our study.

\textbf{Measuring elastic moduli in thermal systems via stress-fluctuations.}
The elastic modulus can be decomposed to the affine and nonaffine components, expressed as
\begin{equation}
    C^{\alpha\beta\kappa\chi}=C_{\rm A}^{\alpha\beta\kappa\chi} -C_{\rm NA}^{\alpha\beta\kappa\chi}.
\end{equation}

The affine modulus is expressed as \cite{mizuno2022computational}:
\begin{align}
    C_{\rm A}^{\alpha\beta\kappa\chi}
    =&C_{\rm B}^{\alpha\beta\kappa\chi} +C_{\rm K}^{\alpha\beta\kappa\chi}+C_{\rm C}^{\alpha\beta\kappa\chi},\label{eq4}\\
    C_{\rm B}^{\alpha\beta\kappa\chi}
    =&\frac{1}{V}\sum_{i<j}(r_{ij}k_{ij}+f_{ij})r_{ij}n_{ij}^{\alpha}n_{ij}^{\beta}n_{ij}^{\kappa}n_{ij}^{\chi},\\
    C_{\rm K}^{\alpha\beta\kappa\chi}
    =&2\rho T(\delta_{\alpha\kappa} \delta_{\beta\chi}+\delta_{\alpha\chi} \delta_{\beta\kappa}),\\
    C_{\rm C}^{\alpha\beta\kappa\chi}
    =&\frac12 (2\langle\sigma_{\alpha\beta}\rangle\delta_{\kappa\chi}-\langle\sigma_{\alpha\kappa}\rangle\delta_{\beta\chi}-\langle\sigma_{\alpha\chi}\rangle\delta_{\beta\kappa}
    \\&-\langle\sigma_{\beta\kappa}\rangle\delta_{\alpha\chi}-\langle\sigma_{\beta\chi}\rangle\delta_{\alpha\kappa}),\label{eq7}\\
    \sigma_{\alpha\beta}=&\rho k_{\rm B}T\delta_{\alpha\beta}+ \frac{1}{V}\sum_{i<j}f_{ij}^{\alpha}r_{ij}^{\beta}, 
\end{align}

where $k_{ij}=\partial^2u_{ij}/\partial r_{ij}^2$ is the spring constant, $f_{ij}=-\partial u_{ij}/\partial r_{ij}$ is the interparticle force, $r_{ij}^{\alpha}=r_{i}^{\alpha}-r_{j}^{\alpha}$ is the position vector, $n_{ij}^{\alpha}=r_{ij}^{\alpha}/r_{ij}$, $\langle \rangle$ denotes the ensemble average, $V$ is the area or volume of the system, $\rho$ is the number density, $\sigma_{\alpha\beta}$ is the stress tensor. Here $C_{\rm B}^{\alpha\beta\kappa\chi}$ is the so-called Born term, $C_{\rm K}^{\alpha\beta\kappa\chi}$ is the kinetic term, and $C_{\rm C}^{\alpha\beta\kappa\chi}$ is the pre-stress term. The kinetic and pre-stress terms are often negligible for low-temperature amorphous solids.

The nonaffine modulus can be expressed as the fluctuation of stress, given by
\begin{equation}
    C_{\rm NA}^{\alpha\beta\kappa\chi}=\frac{V}{T}\left[\langle\sigma_{\alpha\beta}\sigma_{\kappa\chi}\rangle-\langle \sigma_{\alpha\beta}\rangle\langle\sigma_{\kappa\chi}\rangle\right]. \label{eq9}
\end{equation}

\textbf{Measuring elastic moduli in athermal systems via the Hessian matrix.}
In athermal systems, the affine modulus is calculated using the same formula as in the thermal scenario (Eq.~(\ref{eq4})-(\ref{eq7})), excluding the kinetic term. The nonaffine modulus is represented using the Hessian matrix $\mathbf{H}$~\cite{lemaitre06modulus,mizuno2022computational}.

The Hessian matrix $\mathbf{H}$ is defined as
\begin{equation}
    \mathbf{H}=\frac{\partial ^2 U}{\partial \mathbf{r}_i \partial \mathbf{r}_j},
\end{equation}
where $U$ represents the total interaction potential:
\begin{equation}
    U(\{r_{ij}\})=\sum_{i>j}u_{ij}(r_{ij}).
\end{equation}

The off-diagonal terms of $\mathbf{H}$ are given by
\begin{align}
    H_{ij}^{\alpha\beta}&=\frac{\partial ^2 u(r_{ij})}{\partial r_i^\alpha \partial r_j^\beta}\\
    &=-(k_{ij}+\frac{f_{ij}}{r_{ij}})n_{ij}^{\alpha}n_{ij}^{\beta}+\frac{f_{ij}}{r_{ij}}\delta_{\alpha\beta},
\end{align}
where $k_{ij}=\partial^2u_{ij}/\partial r_{ij}^2$ is the spring constant, $f_{ij}=-\partial u_{ij}/\partial r_{ij}$ is the interparticle force, $r_{ij}^{\alpha}=r_{i}^{\alpha}-r_{j}^{\alpha}$ is the position vector, and $r_{ij}$ is the distance between particles $i$ and $j$, $n_{ij}^{\alpha}=r_{ij}^{\alpha}/r_{ij}$.
The diagonal terms of $\mathbf{H}$ are $H_{ii}^{\alpha\beta}=-\sum_{j\neq i}H_{ij}^{\alpha\beta}$.

The total force exerted on individual particles post the affine deformation, known as the affine force field $\mathbf{\Xi}_{\kappa\chi}$, is given by
\begin{equation}
    \mathbf{\Xi}_{\kappa\chi}=\Xi_{i,\kappa\chi}^\alpha=-\sum_{j}(r_{ij}k_{ij}+f_{ij})n_{ij}^{\alpha}n_{ij}^{\kappa}n_{ij}^{\chi}.
\end{equation}

The nonaffine modulus tensor can subsequently be expressed as
\begin{equation}
    C_{\rm NA}^{\alpha\beta\kappa\chi}=\frac{1}{V}\mathbf{\Xi}_{\alpha\beta}\mathbf{H}^{-1}\mathbf{\Xi}_{\kappa\chi}. 
\end{equation}

\textbf{Measuring elastic moduli via oscillatory shear.}
The elastic moduli of a material can be ascertained by employing oscillatory deformation in simulations, a method analogous to the experimental approach of determining a material's viscoelastic response with a rheometer.

The induced strain is represented by a sinusoidal function:
\begin{equation}
\epsilon(t)=\epsilon_0\sin{(\omega t)}
\end{equation}
where $\epsilon_0$ specifies the amplitude, and $\omega$ indicates the frequency of the oscillatory deformation.

The stress responses to this strain can be expressed as follows:
\begin{equation}
    \begin{split}
    \sigma(t)&=\sigma_0\sin{(\omega t+\delta)}+\sigma_{\rm c} \\&= \sigma_0\cos{\delta}\sin{(\omega t)}+\sigma_0\sin{\delta}\cos{(\omega t)}+\sigma_{\rm c}.
    \end{split}
\end{equation}

The storage modulus, represented by the in-phase term $M'=(\sigma_0\cos{\delta})/\epsilon_0$, and the loss modulus, denoted by the out-of-phase term $M''=(\sigma_0\sin{\delta})/\epsilon_0$, can be derived accordingly. The variable $\sigma_{\rm c}$ denotes the initial stress within the material, typically zero for shear stress but finite in the case of pressure.

The complex shear modulus $G'+iG''$ can be obtained by implementing a shear strain $\epsilon_{xy}$ and measuring the correlated shear stress response $\epsilon_{xy}$. Similarly, by applying a volumetric strain $\epsilon_{\rm b}$ and evaluating the corresponding pressure response $p$, the complex bulk modulus $K'+iK''$ can be calculated. 

Initially, we utilize an amplitude sweep to identify the linear-response region. Upon selecting an amplitude within this region, we are able to ascertain the frequency-dependent storage and loss moduli, as illustrated in Fig. \ref{fig:Sfigure3}.

In this context, our primary interest is the storage modulus. Therefore, in the main text, $G$ and $K$ refer exclusively to the storage moduli.

\textbf{Anisotropy parameter of Voronoi cells in glasses.}
In the case of polydisperse or bidisperse mixtures, the radical Voronoi tessellation is employed to acquire the Voronoi cells. The anisotropy of the Voronoi cells is assessed by calculating the ratio of the smallest to largest eigenvalues of the Minkowski tensors~\cite{turk2010disordered}. In this context, the Minkowski tensor $\boldsymbol{\rm W}^{20}_0=\int_K \boldsymbol{r}^2{\rm d}V$ is utilized, representing the inertial tensor of the Voronoi cell. Subsequently, the anisotropy parameter is expressed as $\beta=I_{\min}/I_{\max}$, where $I$ represents the eigenvalue of $\boldsymbol{\rm W}^{20}_0$. This anisotropy parameter $\beta$ quantifies the degree of asymmetry of the Voronoi cells.

\textbf{Orientational order parameter in glasses.}
The generalized orientational order parameter $\Theta$~\cite{tong2018revealing,tong2019structural} measures the extent to which a local packing deviates from the ideal arrangement where neighboring particles can be maximally efficiently packed around the central particle. For a particle $o$ and its adjacent neighbors $i$ and $j$, the angle between $oi$ and $oj$ is denoted as $\theta_{ij}^{(1)}$. Here the neighbors are given by the radical Voronoi tessellation. When particle $o$, $i$, and $j$ are in contact, the angle between $oi$ and $oj$ is denoted as $\theta_{ij}^{(2)}$. The orientational order parameter $\Theta_o$ for particle $o$ is defined as follows:
\begin{equation}
\Theta_o=\frac{1}{N_o}\sum_{\langle ij\rangle}|\theta_{ij}^{(1)}-\theta_{ij}^{(2)}|,
\end{equation}
where $N_o$ represents the number of neighbor pairs, which is equal to the number of neighbors of particle $o$, and the summation includes all adjacent neighbor pairs. Then the orientational order parameter of the whole system is given by $\Theta=\sum_{o=1}^{N}\Theta_o/N$.

\textbf{Loop number for a coarse-grained network in gels.}
To capture the topological feature of the network on the strand scale, we binarize the coarse-grained density field of the largest cluster. The coarse-grained density field is given by $\rho_{\rm cg}(\mathbf{r})=\sum_i\exp{(-|\mathbf{r}-\mathbf{R}_i|^2/2\Delta_{\rm cg}^2)}$, where $\Delta_{\rm cg}=d/2$ is the standard deviation of the Gaussian filter, and $\mathbf{R}_i$ is the position of particle $i$ belonging to the largest cluster. The threshold for the binarisation is set as $1/2$.

This allows us to calculate the Euler number $\chi$ of the coarse-grained network and determine the number of loops (tunnels), using $N_{\rm loop}=N_{\rm obj}+N_{\rm hole}-\chi$. Here, $N_{\rm obj}=1$ is the number of objects, and the number of holes, $N_{\rm hole}$, can be neglected.

\textbf{Relaxation dynamics in glasses and gels.}
In 2D glass formers, we analyse the particle dynamics using cage-relative motion to eliminate the long-wavelength Mermin-Wagner fluctuations~\cite{flenner2015fundamental,shiba2016unveiling,vivek2017long,illing2017mermin}. This motion is defined as:
\begin{equation}
\underline{\boldsymbol{r}}_j(t)=\boldsymbol{r}_j(t)-\frac{1}{n_j}\sum_l \boldsymbol{r}_l(t),
\end{equation}
where $\boldsymbol{r}_j(t)$ represents the position vector of particle $j$, $l$ iterates over its Voronoi neighbors, and $n_j$ denotes the number of radical Voronoi neighbors of particle $j$. The self-intermediate scattering function is defined as:
\begin{equation}
F_{\rm s}(k,t)=\left\langle\frac{1}{N}\sum_j \exp\left(i\boldsymbol{k}\cdot\left[\underline{\boldsymbol{r}}_j(t)-\underline{\boldsymbol{r}}_j(0)\right]\right)\right\rangle,
\end{equation}
where $k=|\boldsymbol{k}|$ is selected as the first peak of the static structure factor. The mean squared displacement is defined as:
\begin{equation}
\langle \Delta r^2(t)\rangle=\left\langle\frac{1}{N}\sum_j \left[\underline{\boldsymbol{r}}_j(t)-\underline{\boldsymbol{r}}_j(0)\right]^2\right\rangle.
\end{equation}

In 3D gels, the original particle positions are used to calculate the $F_{\rm s}(k,t)$ and $\langle \Delta r^2\rangle$.

\setcounter{figure}{0}
\renewcommand{\thefigure}{S\arabic{figure}}

\begin{figure*}
	\centerline{\includegraphics[width = 17 cm]{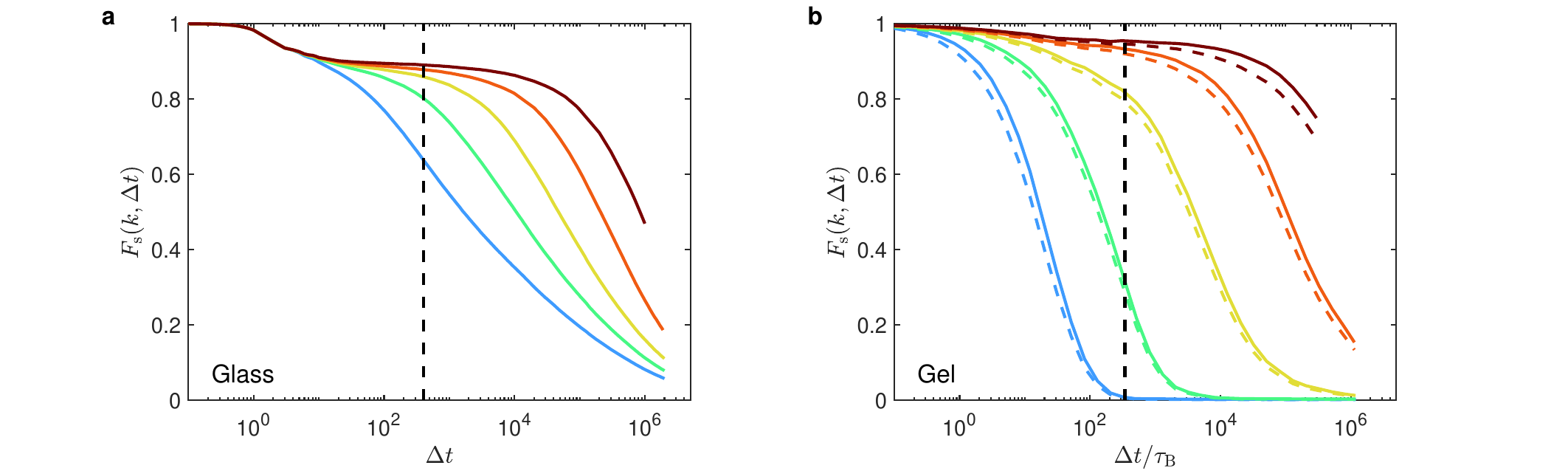}}
	\caption{Self-intermediate scattering function of glasses and gels.
    (a) Cage-relative self-intermediate scattering function $F_{\rm s}(k,\Delta t)$ for glasses at different ages ($t_{\rm w}=10^2,10^3,10^4,10^5,10^6$), depicted from left to right. (b) $F_{\rm s}(k,\Delta t)$ of all particles (dashed lines) and only isostatic particles (solid lines) for gels at different ages ($t_{\rm w}/\tau_{\rm B}=10^2,10^3,10^4,10^5,10^6$), shown from left to right. The vertical dashed line in (a) indicates the time scale used to calculate the stress fluctuations in glasses, while the vertical dashed line in (b) indicates the period of oscillatory deformation used in measuring the moduli of gels.}
	\label{fig:Sfigure1}
\end{figure*}

\begin{figure*}
	\centerline{\includegraphics[width = 17 cm]{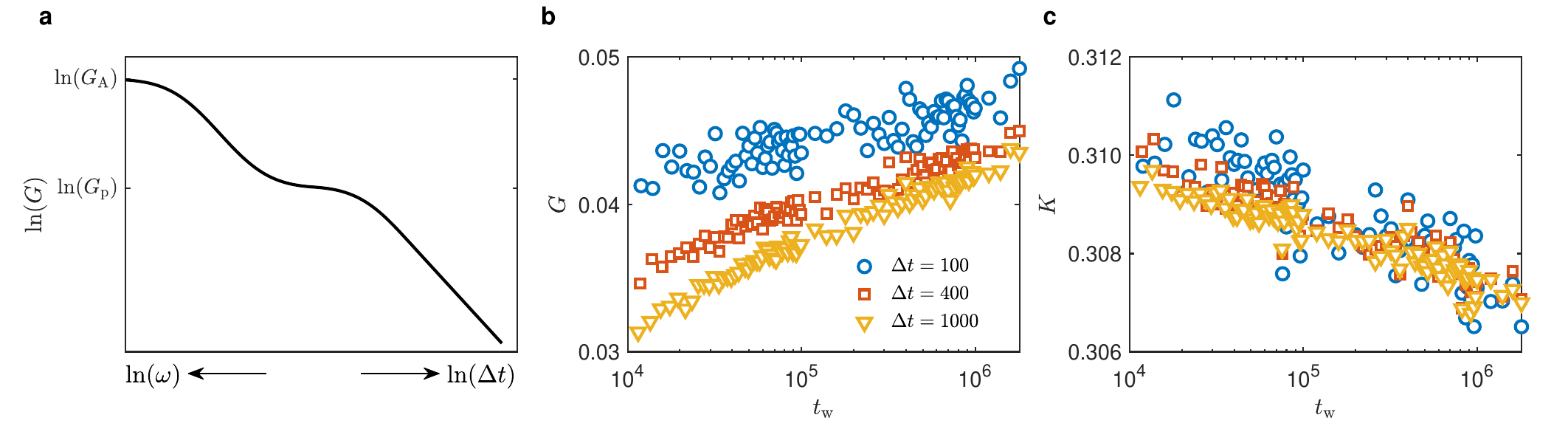}}
	\caption{Age-dependent moduli for different observation times in glasses.
    (a) Schematic representation of the observation-time or frequency-dependent shear modulus. 
    Age-dependent (b) shear modulus and (c) bulk modulus obtained using different statistical times $\Delta t$ for calculating stress fluctuations.}
	\label{fig:Sfigure2}
\end{figure*}

\clearpage 

\begin{figure*}
	\centerline{\includegraphics[width = 17 cm]{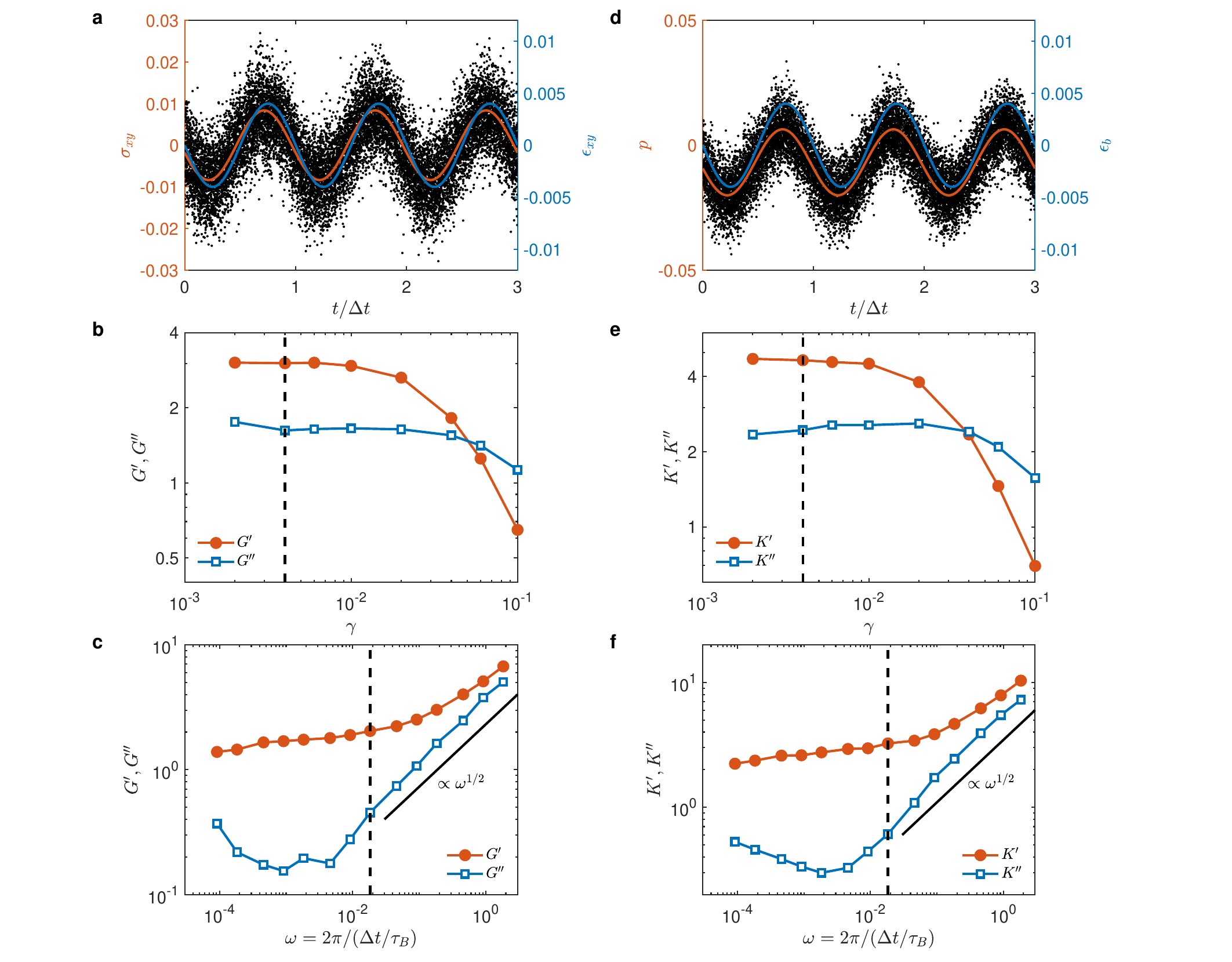}}
    \caption{Amplitude and frequency sweep moduli in gels.
    (a) Shear stress $\sigma_{xy}$ and shear strain $\epsilon_{xy}$ versus time $t/\Delta t$ during oscillatory shear, where $\Delta t$ is the period time. The black points represent individual $\sigma_{xy}$, and the red curve represents the fitting results using $\sigma_{xy}=\sigma_0\sin{(\omega t+\delta)}$.
    (b) Amplitude $\gamma$ sweep of storage shear modulus $G'$ and loss shear modulus $G''$.
    (c) Frequency $\omega=2\pi/(\Delta t/\tau_{\rm B})$ sweep of $G'$ and $G''$.
    (d) Pressure $p$ and volumetric strain $\epsilon_b$ versus $t/\Delta t$ during oscillatory compression. The red curve represents the fitting results using $p=p_0\sin{(\omega t+\delta)}+p_{\rm c}$, where $p_{\rm c}$ is the initial pressure of the system.
    (e) Amplitude $\gamma$ sweep of storage bulk modulus $K'$ and loss bulk modulus $K''$.
    (f) Frequency $\omega=2\pi/(\Delta t/\tau_{\rm B})$ sweep of $K'$ and $K''$. The vertical dashed lines mark the amplitude and frequency used to measure the moduli in Fig. 2. In this work, $G$ and $K$ of gels refer specifically to the storage moduli $G'$ and $K'$.}
	\label{fig:Sfigure3}
\end{figure*}

\begin{figure*}
	\centerline{\includegraphics[width = 17 cm]{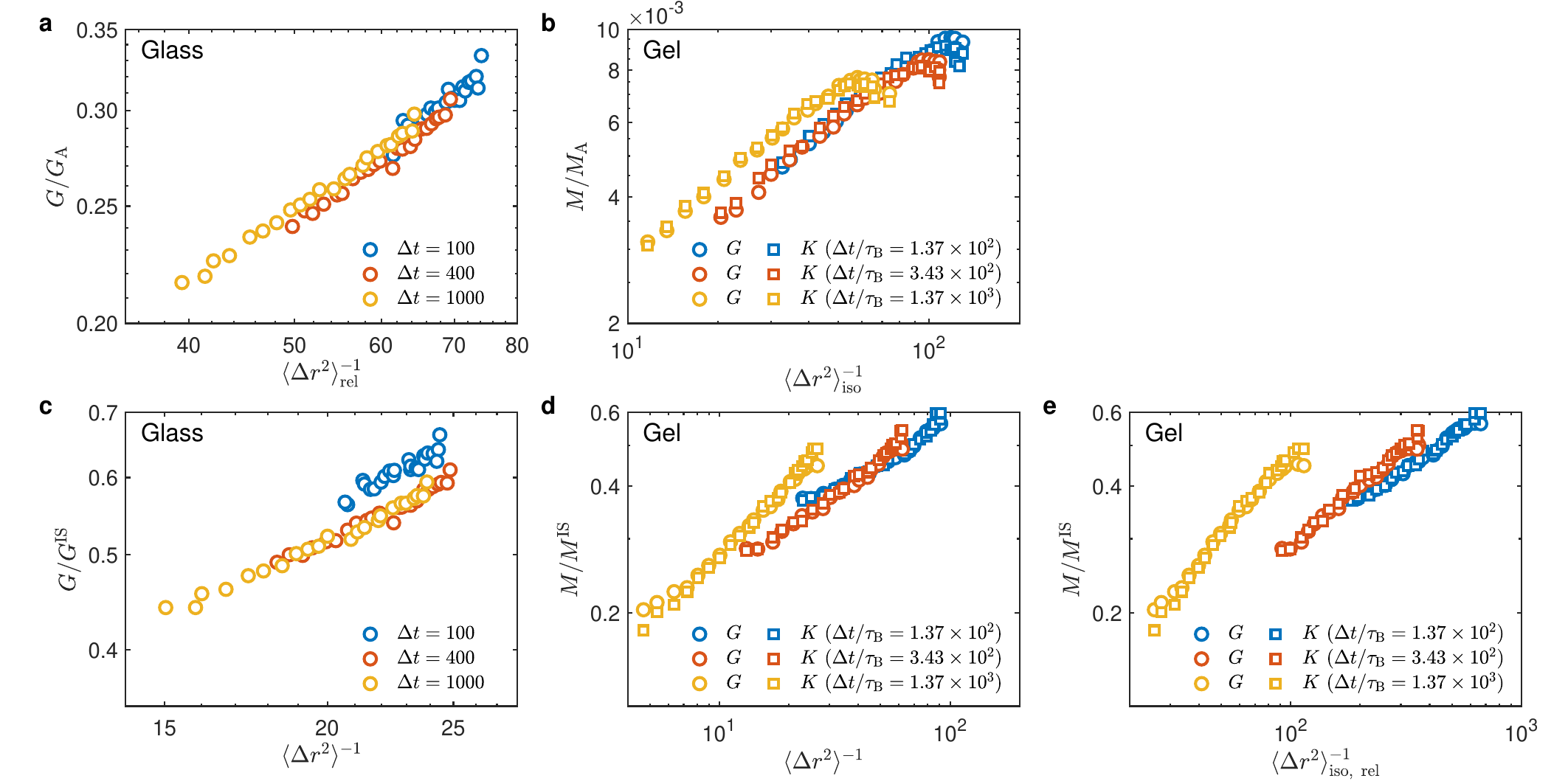}}
	\caption{Rescaled moduli versus mean-squared displacement (MSD) in glasses and gels.
    (a) Shear modulus $G$ rescaled by the affine modulus $G_{\rm A}$ (infinite-frequency modulus) versus the inverse of cage-relative MSD $\langle \Delta r^2\rangle^{-1}_{\rm rel}$ in glasses.
    (b) Shear modulus $G$ rescaled by $G_{\rm A}$ versus the inverse of isostatic MSD $\langle \Delta r^2\rangle^{-1}_{\rm iso}$ in gels.
    (c) Rescaled shear modulus $G/G^{\rm IS}$ versus original MSD $\langle \Delta r^2\rangle^{-1}$ in glasses.
    (d) Rescaled shear modulus $G/G^{\rm IS}$ and bulk modulus $K/K^{\rm IS}$ versus original MSD of all particles $\langle \Delta r^2\rangle^{-1}$ in gels.
    (e) Rescaled shear modulus $G/G^{\rm IS}$ and bulk modulus $K/K^{\rm IS}$ versus cage-relative MSD of isostatic particles $\langle \Delta r^2\rangle^{-1}_{\rm iso,\ rel}$.}
	\label{fig:Sfigure4}
\end{figure*}

\begin{figure*}
	\centerline{\includegraphics[width = 17 cm]{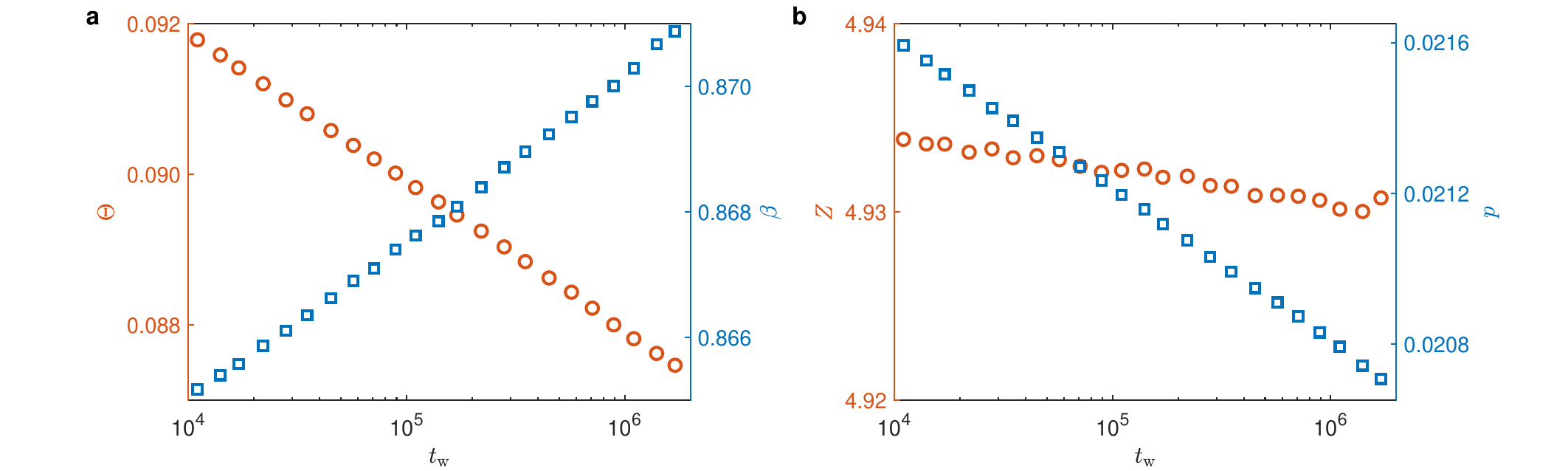}}
	\caption{Age-dependent structure of glasses in inherent states. (a) Orientational order parameter $\Theta$ and cage anisotropy parameter $\beta$. (b) Mean contact number $Z$ and pressure $p$. Note that the scales of all vertical axes are the same as those shown for the corresponding structural parameters in Fig. 4 in the main text.}
	\label{fig:Sfigure5}
\end{figure*}


\begin{thebibliography}{58}%
\makeatletter
\providecommand \@ifxundefined [1]{%
 \@ifx{#1\undefined}
}%
\providecommand \@ifnum [1]{%
 \ifnum #1\expandafter \@firstoftwo
 \else \expandafter \@secondoftwo
 \fi
}%
\providecommand \@ifx [1]{%
 \ifx #1\expandafter \@firstoftwo
 \else \expandafter \@secondoftwo
 \fi
}%
\providecommand \natexlab [1]{#1}%
\providecommand \enquote  [1]{``#1''}%
\providecommand \bibnamefont  [1]{#1}%
\providecommand \bibfnamefont [1]{#1}%
\providecommand \citenamefont [1]{#1}%
\providecommand \href@noop [0]{\@secondoftwo}%
\providecommand \href [0]{\begingroup \@sanitize@url \@href}%
\providecommand \@href[1]{\@@startlink{#1}\@@href}%
\providecommand \@@href[1]{\endgroup#1\@@endlink}%
\providecommand \@sanitize@url [0]{\catcode `\\12\catcode `\$12\catcode
  `\&12\catcode `\#12\catcode `\^12\catcode `\_12\catcode `\%12\relax}%
\providecommand \@@startlink[1]{}%
\providecommand \@@endlink[0]{}%
\providecommand \url  [0]{\begingroup\@sanitize@url \@url }%
\providecommand \@url [1]{\endgroup\@href {#1}{\urlprefix }}%
\providecommand \urlprefix  [0]{URL }%
\providecommand \Eprint [0]{\href }%
\providecommand \doibase [0]{http://dx.doi.org/}%
\providecommand \selectlanguage [0]{\@gobble}%
\providecommand \bibinfo  [0]{\@secondoftwo}%
\providecommand \bibfield  [0]{\@secondoftwo}%
\providecommand \translation [1]{[#1]}%
\providecommand \BibitemOpen [0]{}%
\providecommand \bibitemStop [0]{}%
\providecommand \bibitemNoStop [0]{.\EOS\space}%
\providecommand \EOS [0]{\spacefactor3000\relax}%
\providecommand \BibitemShut  [1]{\csname bibitem#1\endcsname}%
\let\auto@bib@innerbib\@empty
\bibitem [{\citenamefont {Berthier}\ and\ \citenamefont
  {Biroli}(2011)}]{berthier2011theoretical}%
  \BibitemOpen
  \bibfield  {author} {\bibinfo {author} {\bibfnamefont {L.}~\bibnamefont
  {Berthier}}\ and\ \bibinfo {author} {\bibfnamefont {G.}~\bibnamefont
  {Biroli}},\ }\href {\doibase 10.1103/RevModPhys.83.587} {\bibfield  {journal}
  {\bibinfo  {journal} {Reviews of Modern Physics}\ }\textbf {\bibinfo {volume}
  {83}},\ \bibinfo {pages} {587} (\bibinfo {year} {2011})}\BibitemShut
  {NoStop}%
\bibitem [{\citenamefont {Wang}(2012)}]{wang2012elastic}%
  \BibitemOpen
  \bibfield  {author} {\bibinfo {author} {\bibfnamefont {W.~H.}\ \bibnamefont
  {Wang}},\ }\href {\doibase https://doi.org/10.1016/j.pmatsci.2011.07.001}
  {\bibfield  {journal} {\bibinfo  {journal} {Progress in Materials Science}\
  }\textbf {\bibinfo {volume} {57}},\ \bibinfo {pages} {487} (\bibinfo {year}
  {2012})}\BibitemShut {NoStop}%
\bibitem [{\citenamefont {Zaccarelli}(2007)}]{zaccarelli07review}%
  \BibitemOpen
  \bibfield  {author} {\bibinfo {author} {\bibfnamefont {E.}~\bibnamefont
  {Zaccarelli}},\ }\href {\doibase 10.1088/0953-8984/19/32/323101} {\bibfield
  {journal} {\bibinfo  {journal} {Journal of Physics: Condensed Matter}\
  }\textbf {\bibinfo {volume} {19}},\ \bibinfo {pages} {323101} (\bibinfo
  {year} {2007})}\BibitemShut {NoStop}%
\bibitem [{\citenamefont {Royall}\ \emph {et~al.}(2021)\citenamefont {Royall},
  \citenamefont {Faers}, \citenamefont {Fussell},\ and\ \citenamefont
  {Hallett}}]{royall21review}%
  \BibitemOpen
  \bibfield  {author} {\bibinfo {author} {\bibfnamefont {C.~P.}\ \bibnamefont
  {Royall}}, \bibinfo {author} {\bibfnamefont {M.~A.}\ \bibnamefont {Faers}},
  \bibinfo {author} {\bibfnamefont {S.~L.}\ \bibnamefont {Fussell}}, \ and\
  \bibinfo {author} {\bibfnamefont {J.~E.}\ \bibnamefont {Hallett}},\ }\href
  {\doibase 10.1088/1361-648x/ac04cb} {\bibfield  {journal} {\bibinfo
  {journal} {Journal of Physics: Condensed Matter}\ }\textbf {\bibinfo {volume}
  {33}},\ \bibinfo {pages} {453002} (\bibinfo {year} {2021})}\BibitemShut
  {NoStop}%
\bibitem [{\citenamefont {Tanaka}\ \emph {et~al.}(2019)\citenamefont {Tanaka},
  \citenamefont {Tong}, \citenamefont {Shi},\ and\ \citenamefont
  {Russo}}]{tanaka2019revealing}%
  \BibitemOpen
  \bibfield  {author} {\bibinfo {author} {\bibfnamefont {H.}~\bibnamefont
  {Tanaka}}, \bibinfo {author} {\bibfnamefont {H.}~\bibnamefont {Tong}},
  \bibinfo {author} {\bibfnamefont {R.}~\bibnamefont {Shi}}, \ and\ \bibinfo
  {author} {\bibfnamefont {J.}~\bibnamefont {Russo}},\ }\href {\doibase
  10.1038/s42254-019-0053-3} {\bibfield  {journal} {\bibinfo  {journal} {Nature
  Reviews Physics}\ }\textbf {\bibinfo {volume} {1}},\ \bibinfo {pages} {333}
  (\bibinfo {year} {2019})}\BibitemShut {NoStop}%
\bibitem [{\citenamefont {Bonn}\ \emph {et~al.}(2017)\citenamefont {Bonn},
  \citenamefont {Denn}, \citenamefont {Berthier}, \citenamefont {Divoux},\ and\
  \citenamefont {Manneville}}]{bonn2017yield}%
  \BibitemOpen
  \bibfield  {author} {\bibinfo {author} {\bibfnamefont {D.}~\bibnamefont
  {Bonn}}, \bibinfo {author} {\bibfnamefont {M.~M.}\ \bibnamefont {Denn}},
  \bibinfo {author} {\bibfnamefont {L.}~\bibnamefont {Berthier}}, \bibinfo
  {author} {\bibfnamefont {T.}~\bibnamefont {Divoux}}, \ and\ \bibinfo {author}
  {\bibfnamefont {S.}~\bibnamefont {Manneville}},\ }\href {\doibase
  10.1103/RevModPhys.89.035005} {\bibfield  {journal} {\bibinfo  {journal}
  {Reviews of Modern Physics}\ }\textbf {\bibinfo {volume} {89}},\ \bibinfo
  {pages} {035005} (\bibinfo {year} {2017})}\BibitemShut {NoStop}%
\bibitem [{\citenamefont {Lu}\ \emph {et~al.}(2008)\citenamefont {Lu},
  \citenamefont {Zaccarelli}, \citenamefont {Ciulla}, \citenamefont
  {Schofield}, \citenamefont {Sciortino},\ and\ \citenamefont
  {Weitz}}]{lu08gelation}%
  \BibitemOpen
  \bibfield  {author} {\bibinfo {author} {\bibfnamefont {P.~J.}\ \bibnamefont
  {Lu}}, \bibinfo {author} {\bibfnamefont {E.}~\bibnamefont {Zaccarelli}},
  \bibinfo {author} {\bibfnamefont {F.}~\bibnamefont {Ciulla}}, \bibinfo
  {author} {\bibfnamefont {A.~B.}\ \bibnamefont {Schofield}}, \bibinfo {author}
  {\bibfnamefont {F.}~\bibnamefont {Sciortino}}, \ and\ \bibinfo {author}
  {\bibfnamefont {D.~A.}\ \bibnamefont {Weitz}},\ }\href {\doibase
  10.1038/nature06931} {\bibfield  {journal} {\bibinfo  {journal} {Nature}\
  }\textbf {\bibinfo {volume} {453}},\ \bibinfo {pages} {499} (\bibinfo {year}
  {2008})}\BibitemShut {NoStop}%
\bibitem [{\citenamefont {Zia}\ \emph {et~al.}(2014)\citenamefont {Zia},
  \citenamefont {Landrum},\ and\ \citenamefont {Russel}}]{zia2014micro}%
  \BibitemOpen
  \bibfield  {author} {\bibinfo {author} {\bibfnamefont {R.~N.}\ \bibnamefont
  {Zia}}, \bibinfo {author} {\bibfnamefont {B.~J.}\ \bibnamefont {Landrum}}, \
  and\ \bibinfo {author} {\bibfnamefont {W.~B.}\ \bibnamefont {Russel}},\
  }\href {\doibase 10.1122/1.4892115} {\bibfield  {journal} {\bibinfo
  {journal} {Journal of Rheology}\ }\textbf {\bibinfo {volume} {58}},\ \bibinfo
  {pages} {1121} (\bibinfo {year} {2014})}\BibitemShut {NoStop}%
\bibitem [{\citenamefont {Testard}\ \emph {et~al.}(2011)\citenamefont
  {Testard}, \citenamefont {Berthier},\ and\ \citenamefont
  {Kob}}]{testard2011influence}%
  \BibitemOpen
  \bibfield  {author} {\bibinfo {author} {\bibfnamefont {V.}~\bibnamefont
  {Testard}}, \bibinfo {author} {\bibfnamefont {L.}~\bibnamefont {Berthier}}, \
  and\ \bibinfo {author} {\bibfnamefont {W.}~\bibnamefont {Kob}},\ }\href
  {\doibase 10.1103/PhysRevLett.106.125702} {\bibfield  {journal} {\bibinfo
  {journal} {Physical Review Letters}\ }\textbf {\bibinfo {volume} {106}},\
  \bibinfo {pages} {125702} (\bibinfo {year} {2011})}\BibitemShut {NoStop}%
\bibitem [{\citenamefont {Testard}\ \emph {et~al.}(2014)\citenamefont
  {Testard}, \citenamefont {Berthier},\ and\ \citenamefont
  {Kob}}]{testard2014intermittent}%
  \BibitemOpen
  \bibfield  {author} {\bibinfo {author} {\bibfnamefont {V.}~\bibnamefont
  {Testard}}, \bibinfo {author} {\bibfnamefont {L.}~\bibnamefont {Berthier}}, \
  and\ \bibinfo {author} {\bibfnamefont {W.}~\bibnamefont {Kob}},\ }\href
  {\doibase 10.1063/1.4871624} {\bibfield  {journal} {\bibinfo  {journal} {The
  Journal of Chemical Physics}\ }\textbf {\bibinfo {volume} {140}},\ \bibinfo
  {pages} {164502} (\bibinfo {year} {2014})}\BibitemShut {NoStop}%
\bibitem [{\citenamefont {Whitaker}\ \emph {et~al.}(2019)\citenamefont
  {Whitaker}, \citenamefont {Varga}, \citenamefont {Hsiao}, \citenamefont
  {Solomon}, \citenamefont {Swan},\ and\ \citenamefont
  {Furst}}]{whitaker19glassy_cluster}%
  \BibitemOpen
  \bibfield  {author} {\bibinfo {author} {\bibfnamefont {K.~A.}\ \bibnamefont
  {Whitaker}}, \bibinfo {author} {\bibfnamefont {Z.}~\bibnamefont {Varga}},
  \bibinfo {author} {\bibfnamefont {L.~C.}\ \bibnamefont {Hsiao}}, \bibinfo
  {author} {\bibfnamefont {M.~J.}\ \bibnamefont {Solomon}}, \bibinfo {author}
  {\bibfnamefont {J.~W.}\ \bibnamefont {Swan}}, \ and\ \bibinfo {author}
  {\bibfnamefont {E.~M.}\ \bibnamefont {Furst}},\ }\href {\doibase
  10.1038/s41467-019-10039-w} {\bibfield  {journal} {\bibinfo  {journal}
  {Nature Communications}\ }\textbf {\bibinfo {volume} {10}},\ \bibinfo {pages}
  {2237} (\bibinfo {year} {2019})}\BibitemShut {NoStop}%
\bibitem [{\citenamefont {Tsurusawa}\ \emph {et~al.}(2019)\citenamefont
  {Tsurusawa}, \citenamefont {Leocmach}, \citenamefont {Russo},\ and\
  \citenamefont {Tanaka}}]{tsurusawa19isostatic}%
  \BibitemOpen
  \bibfield  {author} {\bibinfo {author} {\bibfnamefont {H.}~\bibnamefont
  {Tsurusawa}}, \bibinfo {author} {\bibfnamefont {M.}~\bibnamefont {Leocmach}},
  \bibinfo {author} {\bibfnamefont {J.}~\bibnamefont {Russo}}, \ and\ \bibinfo
  {author} {\bibfnamefont {H.}~\bibnamefont {Tanaka}},\ }\href {\doibase
  10.1126/sciadv.aav6090} {\bibfield  {journal} {\bibinfo  {journal} {Science
  Advances}\ }\textbf {\bibinfo {volume} {5}},\ \bibinfo {pages} {eaav6090}
  (\bibinfo {year} {2019})}\BibitemShut {NoStop}%
\bibitem [{\citenamefont {Hsiao~Lilian}\ \emph {et~al.}(2012)\citenamefont
  {Hsiao~Lilian}, \citenamefont {Newman~Richmond}, \citenamefont
  {Glotzer~Sharon},\ and\ \citenamefont {Solomon~Michael}}]{hsiao12isostatic}%
  \BibitemOpen
  \bibfield  {author} {\bibinfo {author} {\bibfnamefont {C.}~\bibnamefont
  {Hsiao~Lilian}}, \bibinfo {author} {\bibfnamefont {S.}~\bibnamefont
  {Newman~Richmond}}, \bibinfo {author} {\bibfnamefont {C.}~\bibnamefont
  {Glotzer~Sharon}}, \ and\ \bibinfo {author} {\bibfnamefont {J.}~\bibnamefont
  {Solomon~Michael}},\ }\href {\doibase 10.1073/pnas.1206742109} {\bibfield
  {journal} {\bibinfo  {journal} {Proceedings of the National Academy of
  Sciences}\ }\textbf {\bibinfo {volume} {109}},\ \bibinfo {pages} {16029}
  (\bibinfo {year} {2012})}\BibitemShut {NoStop}%
\bibitem [{\citenamefont {Patrick~Royall}\ \emph {et~al.}(2008)\citenamefont
  {Patrick~Royall}, \citenamefont {Williams}, \citenamefont {Ohtsuka},\ and\
  \citenamefont {Tanaka}}]{royall08arrest}%
  \BibitemOpen
  \bibfield  {author} {\bibinfo {author} {\bibfnamefont {C.}~\bibnamefont
  {Patrick~Royall}}, \bibinfo {author} {\bibfnamefont {S.~R.}\ \bibnamefont
  {Williams}}, \bibinfo {author} {\bibfnamefont {T.}~\bibnamefont {Ohtsuka}}, \
  and\ \bibinfo {author} {\bibfnamefont {H.}~\bibnamefont {Tanaka}},\ }\href
  {\doibase 10.1038/nmat2219} {\bibfield  {journal} {\bibinfo  {journal}
  {Nature Materials}\ }\textbf {\bibinfo {volume} {7}},\ \bibinfo {pages} {556}
  (\bibinfo {year} {2008})}\BibitemShut {NoStop}%
\bibitem [{\citenamefont {Zhang}\ \emph {et~al.}(2019)\citenamefont {Zhang},
  \citenamefont {Zhang}, \citenamefont {Bouzid}, \citenamefont {Rocklin},
  \citenamefont {Del~Gado},\ and\ \citenamefont
  {Mao}}]{zhang19rigidity_percolation}%
  \BibitemOpen
  \bibfield  {author} {\bibinfo {author} {\bibfnamefont {S.}~\bibnamefont
  {Zhang}}, \bibinfo {author} {\bibfnamefont {L.}~\bibnamefont {Zhang}},
  \bibinfo {author} {\bibfnamefont {M.}~\bibnamefont {Bouzid}}, \bibinfo
  {author} {\bibfnamefont {D.~Z.}\ \bibnamefont {Rocklin}}, \bibinfo {author}
  {\bibfnamefont {E.}~\bibnamefont {Del~Gado}}, \ and\ \bibinfo {author}
  {\bibfnamefont {X.}~\bibnamefont {Mao}},\ }\href {\doibase
  10.1103/PhysRevLett.123.058001} {\bibfield  {journal} {\bibinfo  {journal}
  {Physical Review Letters}\ }\textbf {\bibinfo {volume} {123}},\ \bibinfo
  {pages} {058001} (\bibinfo {year} {2019})}\BibitemShut {NoStop}%
\bibitem [{\citenamefont {Tsurusawa}\ and\ \citenamefont
  {Tanaka}(2023)}]{tsurusawa2023hierarchical}%
  \BibitemOpen
  \bibfield  {author} {\bibinfo {author} {\bibfnamefont {H.}~\bibnamefont
  {Tsurusawa}}\ and\ \bibinfo {author} {\bibfnamefont {H.}~\bibnamefont
  {Tanaka}},\ }\href {\doibase 10.1038/s41567-023-02063-x} {\bibfield
  {journal} {\bibinfo  {journal} {Nature Physics}\ } (\bibinfo {year} {2023}),\
  10.1038/s41567-023-02063-x}\BibitemShut {NoStop}%
\bibitem [{\citenamefont {Alexander}(1998)}]{alexander1998amorphous}%
  \BibitemOpen
  \bibfield  {author} {\bibinfo {author} {\bibfnamefont {S.}~\bibnamefont
  {Alexander}},\ }\href {\doibase
  https://doi.org/10.1016/S0370-1573(97)00069-0} {\bibfield  {journal}
  {\bibinfo  {journal} {Physics Reports}\ }\textbf {\bibinfo {volume} {296}},\
  \bibinfo {pages} {65} (\bibinfo {year} {1998})}\BibitemShut {NoStop}%
\bibitem [{\citenamefont {DeGiuli}(2018)}]{degiuli2018field}%
  \BibitemOpen
  \bibfield  {author} {\bibinfo {author} {\bibfnamefont {E.}~\bibnamefont
  {DeGiuli}},\ }\href {\doibase 10.1103/PhysRevLett.121.118001} {\bibfield
  {journal} {\bibinfo  {journal} {Physical Review Letters}\ }\textbf {\bibinfo
  {volume} {121}},\ \bibinfo {pages} {118001} (\bibinfo {year}
  {2018})}\BibitemShut {NoStop}%
\bibitem [{\citenamefont {Lema\^{i}tre}(2018)}]{lemaitre2018stress}%
  \BibitemOpen
  \bibfield  {author} {\bibinfo {author} {\bibfnamefont {A.}~\bibnamefont
  {Lema\^{i}tre}},\ }\href@noop {} {\bibfield  {journal} {\bibinfo  {journal}
  {The Journal of chemical physics}\ }\textbf {\bibinfo {volume} {149}},\
  \bibinfo {pages} {104107} (\bibinfo {year} {2018})}\BibitemShut {NoStop}%
\bibitem [{\citenamefont {Nampoothiri}\ \emph {et~al.}(2020)\citenamefont
  {Nampoothiri}, \citenamefont {Wang}, \citenamefont {Ramola}, \citenamefont
  {Zhang}, \citenamefont {Bhattacharjee},\ and\ \citenamefont
  {Chakraborty}}]{nampoothiri2020emergent}%
  \BibitemOpen
  \bibfield  {author} {\bibinfo {author} {\bibfnamefont {J.~N.}\ \bibnamefont
  {Nampoothiri}}, \bibinfo {author} {\bibfnamefont {Y.}~\bibnamefont {Wang}},
  \bibinfo {author} {\bibfnamefont {K.}~\bibnamefont {Ramola}}, \bibinfo
  {author} {\bibfnamefont {J.}~\bibnamefont {Zhang}}, \bibinfo {author}
  {\bibfnamefont {S.}~\bibnamefont {Bhattacharjee}}, \ and\ \bibinfo {author}
  {\bibfnamefont {B.}~\bibnamefont {Chakraborty}},\ }\href {\doibase
  10.1103/PhysRevLett.125.118002} {\bibfield  {journal} {\bibinfo  {journal}
  {Physical Review Letters}\ }\textbf {\bibinfo {volume} {125}},\ \bibinfo
  {pages} {118002} (\bibinfo {year} {2020})}\BibitemShut {NoStop}%
\bibitem [{\citenamefont {Wang}\ \emph {et~al.}(2020)\citenamefont {Wang},
  \citenamefont {Wang},\ and\ \citenamefont {Zhang}}]{wang2020connecting}%
  \BibitemOpen
  \bibfield  {author} {\bibinfo {author} {\bibfnamefont {Y.}~\bibnamefont
  {Wang}}, \bibinfo {author} {\bibfnamefont {Y.}~\bibnamefont {Wang}}, \ and\
  \bibinfo {author} {\bibfnamefont {J.}~\bibnamefont {Zhang}},\ }\href
  {\doibase 10.1038/s41467-020-18217-x} {\bibfield  {journal} {\bibinfo
  {journal} {Nature Communications}\ }\textbf {\bibinfo {volume} {11}},\
  \bibinfo {pages} {4349} (\bibinfo {year} {2020})}\BibitemShut {NoStop}%
\bibitem [{\citenamefont {Yoshino}\ and\ \citenamefont
  {Mézard}(2010)}]{yoshino2010emergence}%
  \BibitemOpen
  \bibfield  {author} {\bibinfo {author} {\bibfnamefont {H.}~\bibnamefont
  {Yoshino}}\ and\ \bibinfo {author} {\bibfnamefont {M.}~\bibnamefont
  {Mézard}},\ }\href {\doibase 10.1103/PhysRevLett.105.015504} {\bibfield
  {journal} {\bibinfo  {journal} {Physical Review Letters}\ }\textbf {\bibinfo
  {volume} {105}},\ \bibinfo {pages} {015504} (\bibinfo {year}
  {2010})}\BibitemShut {NoStop}%
\bibitem [{\citenamefont {Yoshino}\ and\ \citenamefont
  {Zamponi}(2014)}]{yoshino2014shear}%
  \BibitemOpen
  \bibfield  {author} {\bibinfo {author} {\bibfnamefont {H.}~\bibnamefont
  {Yoshino}}\ and\ \bibinfo {author} {\bibfnamefont {F.}~\bibnamefont
  {Zamponi}},\ }\href {\doibase 10.1103/PhysRevE.90.022302} {\bibfield
  {journal} {\bibinfo  {journal} {Physical Review E}\ }\textbf {\bibinfo
  {volume} {90}},\ \bibinfo {pages} {022302} (\bibinfo {year}
  {2014})}\BibitemShut {NoStop}%
\bibitem [{\citenamefont {Szamel}\ and\ \citenamefont
  {Flenner}(2011)}]{szamel2011emergence}%
  \BibitemOpen
  \bibfield  {author} {\bibinfo {author} {\bibfnamefont {G.}~\bibnamefont
  {Szamel}}\ and\ \bibinfo {author} {\bibfnamefont {E.}~\bibnamefont
  {Flenner}},\ }\href {\doibase 10.1103/PhysRevLett.107.105505} {\bibfield
  {journal} {\bibinfo  {journal} {Physical Review Letters}\ }\textbf {\bibinfo
  {volume} {107}},\ \bibinfo {pages} {105505} (\bibinfo {year}
  {2011})}\BibitemShut {NoStop}%
\bibitem [{\citenamefont {Parisi}\ and\ \citenamefont
  {Zamponi}(2010)}]{parisi2010mean}%
  \BibitemOpen
  \bibfield  {author} {\bibinfo {author} {\bibfnamefont {G.}~\bibnamefont
  {Parisi}}\ and\ \bibinfo {author} {\bibfnamefont {F.}~\bibnamefont
  {Zamponi}},\ }\href {\doibase 10.1103/RevModPhys.82.789} {\bibfield
  {journal} {\bibinfo  {journal} {Reviews of Modern Physics}\ }\textbf
  {\bibinfo {volume} {82}},\ \bibinfo {pages} {789} (\bibinfo {year}
  {2010})}\BibitemShut {NoStop}%
\bibitem [{\citenamefont {Yanagishima}\ \emph {et~al.}(2017)\citenamefont
  {Yanagishima}, \citenamefont {Russo},\ and\ \citenamefont
  {Tanaka}}]{yanagishima2017common}%
  \BibitemOpen
  \bibfield  {author} {\bibinfo {author} {\bibfnamefont {T.}~\bibnamefont
  {Yanagishima}}, \bibinfo {author} {\bibfnamefont {J.}~\bibnamefont {Russo}},
  \ and\ \bibinfo {author} {\bibfnamefont {H.}~\bibnamefont {Tanaka}},\ }\href
  {\doibase 10.1038/ncomms15954} {\bibfield  {journal} {\bibinfo  {journal}
  {Nature Communications}\ }\textbf {\bibinfo {volume} {8}},\ \bibinfo {pages}
  {15954} (\bibinfo {year} {2017})}\BibitemShut {NoStop}%
\bibitem [{\citenamefont {Tong}\ \emph {et~al.}(2020)\citenamefont {Tong},
  \citenamefont {Sengupta},\ and\ \citenamefont {Tanaka}}]{tong20rigidity}%
  \BibitemOpen
  \bibfield  {author} {\bibinfo {author} {\bibfnamefont {H.}~\bibnamefont
  {Tong}}, \bibinfo {author} {\bibfnamefont {S.}~\bibnamefont {Sengupta}}, \
  and\ \bibinfo {author} {\bibfnamefont {H.}~\bibnamefont {Tanaka}},\ }\href
  {\doibase 10.1038/s41467-020-18663-7} {\bibfield  {journal} {\bibinfo
  {journal} {Nature Communications}\ }\textbf {\bibinfo {volume} {11}},\
  \bibinfo {pages} {4863} (\bibinfo {year} {2020})}\BibitemShut {NoStop}%
\bibitem [{\citenamefont {Ding}\ \emph {et~al.}(2016)\citenamefont {Ding},
  \citenamefont {Cheng}, \citenamefont {Sheng}, \citenamefont {Asta},
  \citenamefont {Ritchie},\ and\ \citenamefont {Ma}}]{ding2016universal}%
  \BibitemOpen
  \bibfield  {author} {\bibinfo {author} {\bibfnamefont {J.}~\bibnamefont
  {Ding}}, \bibinfo {author} {\bibfnamefont {Y.-Q.}\ \bibnamefont {Cheng}},
  \bibinfo {author} {\bibfnamefont {H.}~\bibnamefont {Sheng}}, \bibinfo
  {author} {\bibfnamefont {M.}~\bibnamefont {Asta}}, \bibinfo {author}
  {\bibfnamefont {R.~O.}\ \bibnamefont {Ritchie}}, \ and\ \bibinfo {author}
  {\bibfnamefont {E.}~\bibnamefont {Ma}},\ }\href {\doibase
  10.1038/ncomms13733} {\bibfield  {journal} {\bibinfo  {journal} {Nature
  Communications}\ }\textbf {\bibinfo {volume} {7}},\ \bibinfo {pages} {13733}
  (\bibinfo {year} {2016})}\BibitemShut {NoStop}%
\bibitem [{\citenamefont {Saw}\ and\ \citenamefont
  {Harrowell}(2016)}]{saw16modulus_calculation}%
  \BibitemOpen
  \bibfield  {author} {\bibinfo {author} {\bibfnamefont {S.}~\bibnamefont
  {Saw}}\ and\ \bibinfo {author} {\bibfnamefont {P.}~\bibnamefont
  {Harrowell}},\ }\href {\doibase 10.1103/PhysRevLett.116.137801} {\bibfield
  {journal} {\bibinfo  {journal} {Physical Review Letters}\ }\textbf {\bibinfo
  {volume} {116}},\ \bibinfo {pages} {137801} (\bibinfo {year}
  {2016})}\BibitemShut {NoStop}%
\bibitem [{\citenamefont {Greinert}\ \emph {et~al.}(2006)\citenamefont
  {Greinert}, \citenamefont {Wood},\ and\ \citenamefont
  {Bartlett}}]{greinert2006measurement}%
  \BibitemOpen
  \bibfield  {author} {\bibinfo {author} {\bibfnamefont {N.}~\bibnamefont
  {Greinert}}, \bibinfo {author} {\bibfnamefont {T.}~\bibnamefont {Wood}}, \
  and\ \bibinfo {author} {\bibfnamefont {P.}~\bibnamefont {Bartlett}},\ }\href
  {\doibase 10.1103/PhysRevLett.97.265702} {\bibfield  {journal} {\bibinfo
  {journal} {Physical Review Letters}\ }\textbf {\bibinfo {volume} {97}},\
  \bibinfo {pages} {265702} (\bibinfo {year} {2006})}\BibitemShut {NoStop}%
\bibitem [{\citenamefont {C.~K.~Poon}\ \emph {et~al.}(1999)\citenamefont
  {C.~K.~Poon}, \citenamefont {Starrs}, \citenamefont {P.~Meeker},
  \citenamefont {Moussa\''{i}d}, \citenamefont {M.~L.~Evans}, \citenamefont
  {N.~Pusey},\ and\ \citenamefont {M.~Robins}}]{poon1999delayed}%
  \BibitemOpen
  \bibfield  {author} {\bibinfo {author} {\bibfnamefont {W.}~\bibnamefont
  {C.~K.~Poon}}, \bibinfo {author} {\bibfnamefont {L.}~\bibnamefont {Starrs}},
  \bibinfo {author} {\bibfnamefont {S.}~\bibnamefont {P.~Meeker}}, \bibinfo
  {author} {\bibfnamefont {A.}~\bibnamefont {Moussa\''{i}d}}, \bibinfo {author}
  {\bibfnamefont {R.}~\bibnamefont {M.~L.~Evans}}, \bibinfo {author}
  {\bibfnamefont {P.}~\bibnamefont {N.~Pusey}}, \ and\ \bibinfo {author}
  {\bibfnamefont {M.}~\bibnamefont {M.~Robins}},\ }\href {\doibase
  10.1039/A900664H} {\bibfield  {journal} {\bibinfo  {journal} {Faraday
  Discussions}\ }\textbf {\bibinfo {volume} {112}},\ \bibinfo {pages} {143}
  (\bibinfo {year} {1999})}\BibitemShut {NoStop}%
\bibitem [{\citenamefont {Teece}\ \emph {et~al.}(2014)\citenamefont {Teece},
  \citenamefont {Hart}, \citenamefont {Hsu}, \citenamefont {Gilligan},
  \citenamefont {Faers},\ and\ \citenamefont
  {Bartlett}}]{teece14delayed_collapse}%
  \BibitemOpen
  \bibfield  {author} {\bibinfo {author} {\bibfnamefont {L.~J.}\ \bibnamefont
  {Teece}}, \bibinfo {author} {\bibfnamefont {J.~M.}\ \bibnamefont {Hart}},
  \bibinfo {author} {\bibfnamefont {K.~Y.~N.}\ \bibnamefont {Hsu}}, \bibinfo
  {author} {\bibfnamefont {S.}~\bibnamefont {Gilligan}}, \bibinfo {author}
  {\bibfnamefont {M.~A.}\ \bibnamefont {Faers}}, \ and\ \bibinfo {author}
  {\bibfnamefont {P.}~\bibnamefont {Bartlett}},\ }\href {\doibase
  https://doi.org/10.1016/j.colsurfa.2014.03.018} {\bibfield  {journal}
  {\bibinfo  {journal} {Colloids and Surfaces A: Physicochemical and
  Engineering Aspects}\ }\textbf {\bibinfo {volume} {458}},\ \bibinfo {pages}
  {126} (\bibinfo {year} {2014})}\BibitemShut {NoStop}%
\bibitem [{\citenamefont {Bartlett}\ \emph {et~al.}(2012)\citenamefont
  {Bartlett}, \citenamefont {Teece},\ and\ \citenamefont
  {Faers}}]{bartlett12delayed_collapse}%
  \BibitemOpen
  \bibfield  {author} {\bibinfo {author} {\bibfnamefont {P.}~\bibnamefont
  {Bartlett}}, \bibinfo {author} {\bibfnamefont {L.~J.}\ \bibnamefont {Teece}},
  \ and\ \bibinfo {author} {\bibfnamefont {M.~A.}\ \bibnamefont {Faers}},\
  }\href {\doibase 10.1103/PhysRevE.85.021404} {\bibfield  {journal} {\bibinfo
  {journal} {Physical Review E}\ }\textbf {\bibinfo {volume} {85}},\ \bibinfo
  {pages} {021404} (\bibinfo {year} {2012})}\BibitemShut {NoStop}%
\bibitem [{\citenamefont {Kamp}\ and\ \citenamefont
  {Kilfoil}(2009)}]{kamp2009universal}%
  \BibitemOpen
  \bibfield  {author} {\bibinfo {author} {\bibfnamefont {S.~W.}\ \bibnamefont
  {Kamp}}\ and\ \bibinfo {author} {\bibfnamefont {M.~L.}\ \bibnamefont
  {Kilfoil}},\ }\href {\doibase 10.1039/B814975E} {\bibfield  {journal}
  {\bibinfo  {journal} {Soft Matter}\ }\textbf {\bibinfo {volume} {5}},\
  \bibinfo {pages} {2438} (\bibinfo {year} {2009})}\BibitemShut {NoStop}%
\bibitem [{\citenamefont {Clarke}(2021)}]{clarke21peak_modulus}%
  \BibitemOpen
  \bibfield  {author} {\bibinfo {author} {\bibfnamefont {A.}~\bibnamefont
  {Clarke}},\ }\href {\doibase 10.1039/D1SM00816A} {\bibfield  {journal}
  {\bibinfo  {journal} {Soft Matter}\ }\textbf {\bibinfo {volume} {17}},\
  \bibinfo {pages} {7893} (\bibinfo {year} {2021})}\BibitemShut {NoStop}%
\bibitem [{\citenamefont {Fenton}\ \emph {et~al.}(2023)\citenamefont {Fenton},
  \citenamefont {Padmanabhan}, \citenamefont {Ryu}, \citenamefont {Nguyen},
  \citenamefont {Zia},\ and\ \citenamefont {Helgeson}}]{fenton2023minimal}%
  \BibitemOpen
  \bibfield  {author} {\bibinfo {author} {\bibfnamefont {S.~M.}\ \bibnamefont
  {Fenton}}, \bibinfo {author} {\bibfnamefont {P.}~\bibnamefont {Padmanabhan}},
  \bibinfo {author} {\bibfnamefont {B.~K.}\ \bibnamefont {Ryu}}, \bibinfo
  {author} {\bibfnamefont {T.~T.~D.}\ \bibnamefont {Nguyen}}, \bibinfo {author}
  {\bibfnamefont {R.~N.}\ \bibnamefont {Zia}}, \ and\ \bibinfo {author}
  {\bibfnamefont {M.~E.}\ \bibnamefont {Helgeson}},\ }\href {\doibase
  10.1073/pnas.2215922120} {\bibfield  {journal} {\bibinfo  {journal}
  {Proceedings of the National Academy of Sciences}\ }\textbf {\bibinfo
  {volume} {120}},\ \bibinfo {pages} {e2215922120} (\bibinfo {year}
  {2023})}\BibitemShut {NoStop}%
\bibitem [{\citenamefont {Bouzid}\ \emph {et~al.}(2018)\citenamefont {Bouzid},
  \citenamefont {Keshavarz}, \citenamefont {Geri}, \citenamefont {Divoux},
  \citenamefont {Del~Gado},\ and\ \citenamefont
  {McKinley}}]{bouzid2018computing}%
  \BibitemOpen
  \bibfield  {author} {\bibinfo {author} {\bibfnamefont {M.}~\bibnamefont
  {Bouzid}}, \bibinfo {author} {\bibfnamefont {B.}~\bibnamefont {Keshavarz}},
  \bibinfo {author} {\bibfnamefont {M.}~\bibnamefont {Geri}}, \bibinfo {author}
  {\bibfnamefont {T.}~\bibnamefont {Divoux}}, \bibinfo {author} {\bibfnamefont
  {E.}~\bibnamefont {Del~Gado}}, \ and\ \bibinfo {author} {\bibfnamefont
  {G.~H.}\ \bibnamefont {McKinley}},\ }\href {\doibase 10.1122/1.5018715}
  {\bibfield  {journal} {\bibinfo  {journal} {Journal of Rheology}\ }\textbf
  {\bibinfo {volume} {62}},\ \bibinfo {pages} {1037} (\bibinfo {year}
  {2018})}\BibitemShut {NoStop}%
\bibitem [{\citenamefont {Bantawa}\ \emph {et~al.}(2023)\citenamefont
  {Bantawa}, \citenamefont {Keshavarz}, \citenamefont {Geri}, \citenamefont
  {Bouzid}, \citenamefont {Divoux}, \citenamefont {McKinley},\ and\
  \citenamefont {Del~Gado}}]{bantawa2023hidden}%
  \BibitemOpen
  \bibfield  {author} {\bibinfo {author} {\bibfnamefont {M.}~\bibnamefont
  {Bantawa}}, \bibinfo {author} {\bibfnamefont {B.}~\bibnamefont {Keshavarz}},
  \bibinfo {author} {\bibfnamefont {M.}~\bibnamefont {Geri}}, \bibinfo {author}
  {\bibfnamefont {M.}~\bibnamefont {Bouzid}}, \bibinfo {author} {\bibfnamefont
  {T.}~\bibnamefont {Divoux}}, \bibinfo {author} {\bibfnamefont {G.~H.}\
  \bibnamefont {McKinley}}, \ and\ \bibinfo {author} {\bibfnamefont
  {E.}~\bibnamefont {Del~Gado}},\ }\href {\doibase 10.1038/s41567-023-01988-7}
  {\bibfield  {journal} {\bibinfo  {journal} {Nature Physics}\ } (\bibinfo
  {year} {2023}),\ 10.1038/s41567-023-01988-7}\BibitemShut {NoStop}%
\bibitem [{\citenamefont {Tong}\ and\ \citenamefont
  {Tanaka}(2018)}]{tong2018revealing}%
  \BibitemOpen
  \bibfield  {author} {\bibinfo {author} {\bibfnamefont {H.}~\bibnamefont
  {Tong}}\ and\ \bibinfo {author} {\bibfnamefont {H.}~\bibnamefont {Tanaka}},\
  }\href {\doibase 10.1103/PhysRevX.8.011041} {\bibfield  {journal} {\bibinfo
  {journal} {Physical Review X}\ }\textbf {\bibinfo {volume} {8}},\ \bibinfo
  {pages} {011041} (\bibinfo {year} {2018})}\BibitemShut {NoStop}%
\bibitem [{\citenamefont {Tong}\ and\ \citenamefont
  {Tanaka}(2019)}]{tong2019structural}%
  \BibitemOpen
  \bibfield  {author} {\bibinfo {author} {\bibfnamefont {H.}~\bibnamefont
  {Tong}}\ and\ \bibinfo {author} {\bibfnamefont {H.}~\bibnamefont {Tanaka}},\
  }\href {\doibase 10.1038/s41467-019-13606-3} {\bibfield  {journal} {\bibinfo
  {journal} {Nature Communications}\ }\textbf {\bibinfo {volume} {10}},\
  \bibinfo {pages} {5596} (\bibinfo {year} {2019})}\BibitemShut {NoStop}%
\bibitem [{\citenamefont {Griffiths}\ \emph {et~al.}(2017)\citenamefont
  {Griffiths}, \citenamefont {Turci},\ and\ \citenamefont
  {Royall}}]{griffiths17low_density}%
  \BibitemOpen
  \bibfield  {author} {\bibinfo {author} {\bibfnamefont {S.}~\bibnamefont
  {Griffiths}}, \bibinfo {author} {\bibfnamefont {F.}~\bibnamefont {Turci}}, \
  and\ \bibinfo {author} {\bibfnamefont {C.~P.}\ \bibnamefont {Royall}},\
  }\href {\doibase 10.1063/1.4973351} {\bibfield  {journal} {\bibinfo
  {journal} {The Journal of Chemical Physics}\ }\textbf {\bibinfo {volume}
  {146}},\ \bibinfo {pages} {014905} (\bibinfo {year} {2017})}\BibitemShut
  {NoStop}%
\bibitem [{\citenamefont {Tateno}\ \emph {et~al.}(2022)\citenamefont {Tateno},
  \citenamefont {Yanagishima},\ and\ \citenamefont {Tanaka}}]{tateno22tetra}%
  \BibitemOpen
  \bibfield  {author} {\bibinfo {author} {\bibfnamefont {M.}~\bibnamefont
  {Tateno}}, \bibinfo {author} {\bibfnamefont {T.}~\bibnamefont {Yanagishima}},
  \ and\ \bibinfo {author} {\bibfnamefont {H.}~\bibnamefont {Tanaka}},\ }\href
  {\doibase 10.1063/5.0080403} {\bibfield  {journal} {\bibinfo  {journal} {The
  Journal of Chemical Physics}\ }\textbf {\bibinfo {volume} {156}},\ \bibinfo
  {pages} {084904} (\bibinfo {year} {2022})}\BibitemShut {NoStop}%
\bibitem [{\citenamefont {Widmer-Cooper}\ and\ \citenamefont
  {Harrowell}(2006{\natexlab{a}})}]{widmer2006free}%
  \BibitemOpen
  \bibfield  {author} {\bibinfo {author} {\bibfnamefont {A.}~\bibnamefont
  {Widmer-Cooper}}\ and\ \bibinfo {author} {\bibfnamefont {P.}~\bibnamefont
  {Harrowell}},\ }\href {\doibase
  https://doi.org/10.1016/j.jnoncrysol.2006.01.136} {\bibfield  {journal}
  {\bibinfo  {journal} {Journal of Non-Crystalline Solids}\ }\textbf {\bibinfo
  {volume} {352}},\ \bibinfo {pages} {5098} (\bibinfo {year}
  {2006}{\natexlab{a}})}\BibitemShut {NoStop}%
\bibitem [{\citenamefont {Widmer-Cooper}\ and\ \citenamefont
  {Harrowell}(2006{\natexlab{b}})}]{widmer2006predicting}%
  \BibitemOpen
  \bibfield  {author} {\bibinfo {author} {\bibfnamefont {A.}~\bibnamefont
  {Widmer-Cooper}}\ and\ \bibinfo {author} {\bibfnamefont {P.}~\bibnamefont
  {Harrowell}},\ }\href {\doibase 10.1103/PhysRevLett.96.185701} {\bibfield
  {journal} {\bibinfo  {journal} {Phys Rev Lett}\ }\textbf {\bibinfo {volume}
  {96}},\ \bibinfo {pages} {185701} (\bibinfo {year}
  {2006}{\natexlab{b}})}\BibitemShut {NoStop}%
\bibitem [{\citenamefont {Flenner}\ and\ \citenamefont
  {Szamel}(2015)}]{flenner2015fundamental}%
  \BibitemOpen
  \bibfield  {author} {\bibinfo {author} {\bibfnamefont {E.}~\bibnamefont
  {Flenner}}\ and\ \bibinfo {author} {\bibfnamefont {G.}~\bibnamefont
  {Szamel}},\ }\href {\doibase 10.1038/ncomms8392} {\bibfield  {journal}
  {\bibinfo  {journal} {Nature Communications}\ }\textbf {\bibinfo {volume}
  {6}},\ \bibinfo {pages} {7392} (\bibinfo {year} {2015})}\BibitemShut
  {NoStop}%
\bibitem [{\citenamefont {Shiba}\ \emph {et~al.}(2016)\citenamefont {Shiba},
  \citenamefont {Yamada}, \citenamefont {Kawasaki},\ and\ \citenamefont
  {Kim}}]{shiba2016unveiling}%
  \BibitemOpen
  \bibfield  {author} {\bibinfo {author} {\bibfnamefont {H.}~\bibnamefont
  {Shiba}}, \bibinfo {author} {\bibfnamefont {Y.}~\bibnamefont {Yamada}},
  \bibinfo {author} {\bibfnamefont {T.}~\bibnamefont {Kawasaki}}, \ and\
  \bibinfo {author} {\bibfnamefont {K.}~\bibnamefont {Kim}},\ }\href {\doibase
  10.1103/PhysRevLett.117.245701} {\bibfield  {journal} {\bibinfo  {journal}
  {Physical Review Letters}\ }\textbf {\bibinfo {volume} {117}},\ \bibinfo
  {pages} {245701} (\bibinfo {year} {2016})}\BibitemShut {NoStop}%
\bibitem [{\citenamefont {Vivek}\ \emph {et~al.}(2017)\citenamefont {Vivek},
  \citenamefont {Kelleher}, \citenamefont {Chaikin},\ and\ \citenamefont
  {Weeks}}]{vivek2017long}%
  \BibitemOpen
  \bibfield  {author} {\bibinfo {author} {\bibfnamefont {S.}~\bibnamefont
  {Vivek}}, \bibinfo {author} {\bibfnamefont {C.~P.}\ \bibnamefont {Kelleher}},
  \bibinfo {author} {\bibfnamefont {P.~M.}\ \bibnamefont {Chaikin}}, \ and\
  \bibinfo {author} {\bibfnamefont {E.~R.}\ \bibnamefont {Weeks}},\ }\href
  {\doibase 10.1073/pnas.1607226113} {\bibfield  {journal} {\bibinfo  {journal}
  {Proceedings of the National Academy of Sciences}\ }\textbf {\bibinfo
  {volume} {114}},\ \bibinfo {pages} {1850} (\bibinfo {year}
  {2017})}\BibitemShut {NoStop}%
\bibitem [{\citenamefont {Illing}\ \emph {et~al.}(2017)\citenamefont {Illing},
  \citenamefont {Fritschi}, \citenamefont {Kaiser}, \citenamefont {Klix},
  \citenamefont {Maret},\ and\ \citenamefont {Keim}}]{illing2017mermin}%
  \BibitemOpen
  \bibfield  {author} {\bibinfo {author} {\bibfnamefont {B.}~\bibnamefont
  {Illing}}, \bibinfo {author} {\bibfnamefont {S.}~\bibnamefont {Fritschi}},
  \bibinfo {author} {\bibfnamefont {H.}~\bibnamefont {Kaiser}}, \bibinfo
  {author} {\bibfnamefont {C.~L.}\ \bibnamefont {Klix}}, \bibinfo {author}
  {\bibfnamefont {G.}~\bibnamefont {Maret}}, \ and\ \bibinfo {author}
  {\bibfnamefont {P.}~\bibnamefont {Keim}},\ }\href {\doibase
  10.1073/pnas.1612964114} {\bibfield  {journal} {\bibinfo  {journal}
  {Proceedings of the National Academy of Sciences}\ }\textbf {\bibinfo
  {volume} {114}},\ \bibinfo {pages} {1856} (\bibinfo {year}
  {2017})}\BibitemShut {NoStop}%
\bibitem [{\citenamefont {Mizuno}\ and\ \citenamefont
  {Ikeda}(2022)}]{mizuno2022computational}%
  \BibitemOpen
  \bibfield  {author} {\bibinfo {author} {\bibfnamefont {H.}~\bibnamefont
  {Mizuno}}\ and\ \bibinfo {author} {\bibfnamefont {A.}~\bibnamefont {Ikeda}},\
  }\enquote {\bibinfo {title} {Computational simulations of the vibrational
  properties of glasses},}\ in\ \href {\doibase doi:10.1142/9781800612587_0010
  10.1142/9781800612587_0010} {\emph {\bibinfo {booktitle} {Low-Temperature
  Thermal and Vibrational Properties of Disordered Solids}}}\ (\bibinfo
  {publisher} {WORLD SCIENTIFIC (EUROPE)},\ \bibinfo {year} {2022})\ pp.\
  \bibinfo {pages} {375--433}\BibitemShut {NoStop}%
\bibitem [{\citenamefont {Kriuchevskyi}\ \emph {et~al.}(2017)\citenamefont
  {Kriuchevskyi}, \citenamefont {Wittmer}, \citenamefont {Meyer},\ and\
  \citenamefont {Baschnagel}}]{kriuchevskyi2017shear}%
  \BibitemOpen
  \bibfield  {author} {\bibinfo {author} {\bibfnamefont {I.}~\bibnamefont
  {Kriuchevskyi}}, \bibinfo {author} {\bibfnamefont {J.~P.}\ \bibnamefont
  {Wittmer}}, \bibinfo {author} {\bibfnamefont {H.}~\bibnamefont {Meyer}}, \
  and\ \bibinfo {author} {\bibfnamefont {J.}~\bibnamefont {Baschnagel}},\
  }\href {\doibase 10.1103/PhysRevLett.119.147802} {\bibfield  {journal}
  {\bibinfo  {journal} {Physical Review Letters}\ }\textbf {\bibinfo {volume}
  {119}},\ \bibinfo {pages} {147802} (\bibinfo {year} {2017})}\BibitemShut
  {NoStop}%
\bibitem [{\citenamefont {Lema\^{i}tre}\ and\ \citenamefont
  {Maloney}(2006)}]{lemaitre06modulus}%
  \BibitemOpen
  \bibfield  {author} {\bibinfo {author} {\bibfnamefont {A.}~\bibnamefont
  {Lema\^{i}tre}}\ and\ \bibinfo {author} {\bibfnamefont {C.}~\bibnamefont
  {Maloney}},\ }\href@noop {} {\bibfield  {journal} {\bibinfo  {journal}
  {Journal of statistical physics}\ }\textbf {\bibinfo {volume} {123}},\
  \bibinfo {pages} {415} (\bibinfo {year} {2006})}\BibitemShut {NoStop}%
\bibitem [{\citenamefont {Rocklin}\ \emph {et~al.}(2021)\citenamefont
  {Rocklin}, \citenamefont {Hsiao}, \citenamefont {Szakasits}, \citenamefont
  {Solomon},\ and\ \citenamefont {Mao}}]{rocklin21elasticity}%
  \BibitemOpen
  \bibfield  {author} {\bibinfo {author} {\bibfnamefont {D.~Z.}\ \bibnamefont
  {Rocklin}}, \bibinfo {author} {\bibfnamefont {L.}~\bibnamefont {Hsiao}},
  \bibinfo {author} {\bibfnamefont {M.}~\bibnamefont {Szakasits}}, \bibinfo
  {author} {\bibfnamefont {M.~J.}\ \bibnamefont {Solomon}}, \ and\ \bibinfo
  {author} {\bibfnamefont {X.}~\bibnamefont {Mao}},\ }\href {\doibase
  10.1039/D0SM00053A} {\bibfield  {journal} {\bibinfo  {journal} {Soft Matter}\
  }\textbf {\bibinfo {volume} {17}},\ \bibinfo {pages} {6929} (\bibinfo {year}
  {2021})}\BibitemShut {NoStop}%
\bibitem [{\citenamefont {Bonacci}\ \emph {et~al.}(2020)\citenamefont
  {Bonacci}, \citenamefont {Chateau}, \citenamefont {Furst}, \citenamefont
  {Fusier}, \citenamefont {Goyon},\ and\ \citenamefont
  {Lema\^{i}tre}}]{bonacci2020contact}%
  \BibitemOpen
  \bibfield  {author} {\bibinfo {author} {\bibfnamefont {F.}~\bibnamefont
  {Bonacci}}, \bibinfo {author} {\bibfnamefont {X.}~\bibnamefont {Chateau}},
  \bibinfo {author} {\bibfnamefont {E.~M.}\ \bibnamefont {Furst}}, \bibinfo
  {author} {\bibfnamefont {J.}~\bibnamefont {Fusier}}, \bibinfo {author}
  {\bibfnamefont {J.}~\bibnamefont {Goyon}}, \ and\ \bibinfo {author}
  {\bibfnamefont {A.}~\bibnamefont {Lema\^{i}tre}},\ }\href {\doibase
  10.1038/s41563-020-0624-9} {\bibfield  {journal} {\bibinfo  {journal} {Nature
  Materials}\ }\textbf {\bibinfo {volume} {19}},\ \bibinfo {pages} {775}
  (\bibinfo {year} {2020})}\BibitemShut {NoStop}%
\bibitem [{\citenamefont {Zhang}\ \emph {et~al.}(2023)\citenamefont {Zhang},
  \citenamefont {Liu},\ and\ \citenamefont {Han}}]{zhang2023anisotropic}%
  \BibitemOpen
  \bibfield  {author} {\bibinfo {author} {\bibfnamefont {H.}~\bibnamefont
  {Zhang}}, \bibinfo {author} {\bibfnamefont {F.}~\bibnamefont {Liu}}, \ and\
  \bibinfo {author} {\bibfnamefont {Y.}~\bibnamefont {Han}},\ }\href@noop {}
  {\bibfield  {journal} {\bibinfo  {journal} {arXiv preprint arXiv:2305.04179}\
  } (\bibinfo {year} {2023})}\BibitemShut {NoStop}%
\bibitem [{\citenamefont {O'Hern}\ \emph {et~al.}(2003)\citenamefont {O'Hern},
  \citenamefont {Silbert}, \citenamefont {Liu},\ and\ \citenamefont
  {Nagel}}]{ohern03jamming}%
  \BibitemOpen
  \bibfield  {author} {\bibinfo {author} {\bibfnamefont {C.~S.}\ \bibnamefont
  {O'Hern}}, \bibinfo {author} {\bibfnamefont {L.~E.}\ \bibnamefont {Silbert}},
  \bibinfo {author} {\bibfnamefont {A.~J.}\ \bibnamefont {Liu}}, \ and\
  \bibinfo {author} {\bibfnamefont {S.~R.}\ \bibnamefont {Nagel}},\ }\href@noop
  {} {\bibfield  {journal} {\bibinfo  {journal} {Physical Review E}\ }\textbf
  {\bibinfo {volume} {68}},\ \bibinfo {pages} {011306} (\bibinfo {year}
  {2003})}\BibitemShut {NoStop}%
\bibitem [{\citenamefont {Cheng}\ and\ \citenamefont
  {Ma}(2009)}]{cheng2009configurational}%
  \BibitemOpen
  \bibfield  {author} {\bibinfo {author} {\bibfnamefont {Y.~Q.}\ \bibnamefont
  {Cheng}}\ and\ \bibinfo {author} {\bibfnamefont {E.}~\bibnamefont {Ma}},\
  }\href {\doibase 10.1103/PhysRevB.80.064104} {\bibfield  {journal} {\bibinfo
  {journal} {Physical Review B}\ }\textbf {\bibinfo {volume} {80}},\ \bibinfo
  {pages} {064104} (\bibinfo {year} {2009})}\BibitemShut {NoStop}%
\bibitem [{\citenamefont {Tateno}\ and\ \citenamefont
  {Tanaka}(2019)}]{tateno19_fpd}%
  \BibitemOpen
  \bibfield  {author} {\bibinfo {author} {\bibfnamefont {M.}~\bibnamefont
  {Tateno}}\ and\ \bibinfo {author} {\bibfnamefont {H.}~\bibnamefont
  {Tanaka}},\ }\href {\doibase 10.1038/s41524-019-0178-z} {\bibfield  {journal}
  {\bibinfo  {journal} {npj Computational Materials}\ }\textbf {\bibinfo
  {volume} {5}},\ \bibinfo {pages} {40} (\bibinfo {year} {2019})}\BibitemShut
  {NoStop}%
\bibitem [{\citenamefont {Schr\"{o}der-Turk}\ \emph {et~al.}(2010)\citenamefont
  {Schr\"{o}der-Turk}, \citenamefont {Mickel}, \citenamefont {Schröter},
  \citenamefont {Delaney}, \citenamefont {Saadatfar}, \citenamefont {Senden},
  \citenamefont {Mecke},\ and\ \citenamefont {Aste}}]{turk2010disordered}%
  \BibitemOpen
  \bibfield  {author} {\bibinfo {author} {\bibfnamefont {G.~E.}\ \bibnamefont
  {Schr\"{o}der-Turk}}, \bibinfo {author} {\bibfnamefont {W.}~\bibnamefont
  {Mickel}}, \bibinfo {author} {\bibfnamefont {M.}~\bibnamefont {Schröter}},
  \bibinfo {author} {\bibfnamefont {G.~W.}\ \bibnamefont {Delaney}}, \bibinfo
  {author} {\bibfnamefont {M.}~\bibnamefont {Saadatfar}}, \bibinfo {author}
  {\bibfnamefont {T.~J.}\ \bibnamefont {Senden}}, \bibinfo {author}
  {\bibfnamefont {K.}~\bibnamefont {Mecke}}, \ and\ \bibinfo {author}
  {\bibfnamefont {T.}~\bibnamefont {Aste}},\ }\href {\doibase
  10.1209/0295-5075/90/34001} {\bibfield  {journal} {\bibinfo  {journal}
  {Europhysics Letters}\ }\textbf {\bibinfo {volume} {90}},\ \bibinfo {pages}
  {34001} (\bibinfo {year} {2010})}\BibitemShut {NoStop}%
\end{thebibliography}
\end{document}